\documentclass[journal]{IEEEtran}

\usepackage{balance}
\usepackage[utf8]{inputenc}
\usepackage{graphicx}
\usepackage{pifont}    
\usepackage{hyperref}  

\usepackage[figuresright]{rotating}

\usepackage{soul}
\usepackage[caption=false,font=footnotesize]{subfig}
\usepackage{epstopdf}
\graphicspath{{figures/}}

\usepackage{amssymb,amsmath}

\usepackage{breqn}              
\usepackage{booktabs}
\usepackage{multirow}
\usepackage[official]{eurosym} 
\usepackage[table]{xcolor}
\usepackage{lipsum} 

\usepackage{enumitem}
\usepackage{upgreek}
\usepackage{textcomp}

\usepackage{listings}
\usepackage{backnaur}

\newcommand{\figref}[1]{Fig.~\ref{fig:#1}}
\newcommand{\figsref}[2]{Figs.~\ref{fig:#1}~and~\ref{fig:#2}}
\newcommand{\figtoref}[2]{Figs.~\ref{fig:#1}~through~\ref{fig:#2}}

\newcommand{\tblref}[1]{Table~\ref{tbl:#1}}

\newcommand{\eref}[1]{Eq.~\ref{eqn:#1}}
\newcommand{\esref}[2]{Eqs.~\ref{eqn:#1}~and~\ref{eqn:#2}}

\newcommand{\secref}[1]{Section~\ref{sec:#1}}
\newcommand{\subsec}[1]{Section~\ref{subsec:#1}}
\newcommand{\subsubsec}[1]{Section~\ref{subsubsec:#1}}

\newcommand{\etal}{\mbox{\emph{et al.}}}

\newcommand{\ie}{\mbox{\emph{i.e.,}}}
\newcommand{\eg}{\mbox{\emph{e.g.,}}}

\newcommand{\etc}{\mbox{\emph{etc}}}
\newcommand{\ignore}[1]{} 

\usepackage{amsthm}
\usepackage{amsmath}
\theoremstyle{plain}
\newtheorem{thm}{Theorem}
\theoremstyle{definition}
\newtheorem{prob}[thm]{\emph{Problem}}

\emergencystretch=\maxdimen
\begin{document}


\title{Toward Ultra-Low-Power Remote Health Monitoring: An Optimal and
  Adaptive Compressed Sensing Framework for Activity Recognition}

%

\author{Josué~Pagán,~\IEEEmembership{Student Member,~IEEE,}
  Ramin~Fallahzadeh,~\IEEEmembership{Student Member,~IEEE,}
  Mahdi~Pedram,~\IEEEmembership{Student Member,~IEEE,}
  José~L.~Risco-Martín, José~M.~Moya, José~L.~Ayala,~\IEEEmembership{Senior
    Member,~IEEE,}, and~Hassan~Ghasemzadeh,~\IEEEmembership{Senior
    Member,~IEEE}
\thanks{J. Pagán, J. L. Risco-Martín and J. L. Ayala are with the
  Department of Computer Architecture and Automation, Universidad
  Complutense de Madrid, 28040 Madrid, Spain. R. Fallahzadeh,
  M. Pedram, and H. Ghasemzadeh are with the School of Electrical
  Engineering and Computer Science, Washington State University,
  Pullman, WA 99164, USA. J. M. Moya is with the Universidad
  Politécnica de Madrid, 28040 Madrid, Spain. J. Pagán, J. L.
  Risco-Martín, J. M. Moya and J. L. Ayala are with the Center for
  Computational Simulation, Universidad Politécnica de Madrid, Campus
  de Montegancedo, Boadilla del Monte, 28660 Madrid, Spain; (e-mail:
  {jpagan,jlrisco,jayala}@ucm.es; {rfallahz, mahdi.pedram,
    hassan}@eecs.wsu.edu; josem@die.upm.es).}  }

\markboth{IEEE TRANSACTIONS ON MOBILE COMPUTING,~VOL.~0, N0.~0, SEPTEMBER~2017}%
{Shell \MakeLowercase{\textit{et al.}}: Bare Demo of IEEEtran.cls for IEEE Journals}

%



\maketitle

\begin{abstract}
  Activity recognition, as an important component of behavioral
monitoring and intervention, has attracted enormous attention,
especially in Mobile Cloud Computing (MCC) and Remote Health
Monitoring (RHM) paradigms. While recently resource constrained
wearable devices have been gaining popularity, their battery life is
limited and constrained by the frequent wireless transmission of data
to more computationally powerful back-ends. This paper proposes an
ultra-low power activity recognition system using a novel adaptive
compressed sensing technique that aims to minimize transmission
costs. Coarse-grained on-body sensor localization and unsupervised
clustering modules are devised to autonomously reconfigure the
compressed sensing module for further power saving. We perform a
thorough heuristic optimization using Grammatical Evolution (GE) to
ensure minimal computation overhead of the proposed methodology. Our
evaluation on a real-world dataset and a low power wearable sensing
node demonstrates that our approach can reduce the energy consumption
of the wireless data transmission up to $81.2\%$ and $61.5\%$, with up
to $60.6\%$ and $35.0\%$ overall power savings in comparison with
baseline and a naive state-of-the-art approaches, respectively. These
solutions lead to an average activity recognition accuracy of
$89.0\%$---only $4.8\%$ less than the baseline accuracy---while having
a negligible energy overhead of on-node computation.

\end{abstract}

\begin{IEEEkeywords}
Compressed sensing, activity recognition, feature selection, energy
efficiency, ultra-low power, optimization, adaptive.
\end{IEEEkeywords}

%
\IEEEpeerreviewmaketitle

\section{Introduction}
%
%
%
%
\label{sec:introduction}
The Internet of Things (IoT) brings new opportunities for health care
in unobtrusive monitoring scenarios (eHealth), for proactive personal
eHealth control~\cite{zheng2014unobtrusive}, sports training
applications~\cite{kugler2011mobile}, health-related timely
interventions~\cite{alemdar2010wireless},~\etc. Among all these
applications, the core functionality is activity recognition, which is
a key enabler in context aware systems. Activity recognition has
numerous health-related applications such as prevention, diagnosis,
and monitoring of several illnesses related to functional
impairment~\cite{steele2000quantitating} or Parkinson
disease~\cite{klucken2013unbiased}. In addition, it plays a crucial
role in in Ambient Assisted Living tools designed for
elders~\cite{rashidi2013survey}.

Dozens of private companies have reached the market of wearables and
activity recognition. Smartwatches like Apple Watch, sport chest-bands
or wristbands such as Microsoft Band, are just some examples. New
technologies in wearable devices provide more efficient processor
units to compute even more complex activity recognition
algorithms. The Snapdragon 400 processor is an example of a high
performance processor used in many of state-of-the-art
smartwatches. Low performance microcontrollers (MCUs) can be found in
many different wearable devices, as the 16-bit MSP430 MCU---that
controls the open-source hardware platform Nike Fuelband---or the
8-bit MCUs of Sillicon-Labs for wearables.

Many current wearable activity recognition technologies require that
    end-users follow certain protocols while being monitored using
    sensors that are coupled with the human body. For example, they
    must carry a wearable node at certain body location all the time;
    otherwise, physical activity measurements (\eg~type and intensity
    of activities/movements) will be
    inaccurate~\cite{saeedi2014autolocate}. One
    study~\cite{saeedi2014toward}, showed that in absence of automatic
    sensor localization, the accuracy of an activity recognition
    algorithm drops to $33.6\%$ from $98.4\%$. This means the signal
    attributes of wearables are tightly coupled with their associated
    on-body node location. Further more limiting the wearable on-body
    location, imposes user discomfort and limits the performance
    capacity of wearables. This issue has many real-world
    implications. For example, while one person may prefer to use
    his/her smartphone in his pocket, another person may prefer to
    carry the smartphone in a purse or a belt-clip. While one person
    may prefer to use a wrist-band sensor on his right wrist, another
    person may prefer a left wrist setting. A new trend is wearables
    allows users to wear wearables in more than one predefined on-body
    location. The emerging clip-on fitness trackers such as \lq Misfit
    Shine 2\rq~and \lq Fitbit Zip\rq~are examples of this new trend.

Using the Snapdragon 400 processor,
Bhattacharya~\etal~\cite{bhattacharya2016smart} present a methodology
for activity recognition using deep learning. These algorithms need
computational resources that might not be present in more simple
devices. The authors claimed that their system could be used and
implemented on off-the-shelf devices, however, they did not provide
experimental proof of low power consumption. Maurer~\etal~developed a
framework called \emph{eWatch}~\cite{maurer2006location}, which uses
several sensors for location and activity recognition but their main
goal was not on power efficiency. Therefore, it is unlikely that their
system can be utilized in every day settings because of limited
battery lifetime. In~\cite{patterson2005fine},
Patterson~\etal~described a context-aware graphical model to perform a
fine-grained activity recognition using Hidden Markov Models. The
system considers a different sensor modality, which uses tags in
objects and a RFID antenna. While the methodology is interesting, its
application is limited to smart environments such as smart homes or
smart nursing facilities. The authors in~\cite{choudhury2008mobile},
presented a Mobile Sensing Platform for daily activity
recognition. The system uses a large number of sensors, which results
in reduced battery life. Thus, due to the aforementioned limitations,
we will use low performance MCUs to deal with the activity recognition
problem, because they are cheap, highly integrated, and low power.


Physical activity recognition, as one of the most popular applications
of human-centered IoT, is the specific target of this study. However,
this methodology can be potentially extended to other applications of
IoT where power efficiency is an obstacle, such as most of eHealth
applications involving continuous monitoring.  For example, monitoring
Heart Rate Variability (HRV) for detection of cardiac failures, or
photoplethysmography (PPG) for information extraction such as blood
pressure or oxygen saturation.

In this paradigm we can distinguish, at least, two scenarios: i) a
sensing node with some knowledge (\ie~simple feature extractions) that
transmits data to a gateway platform (e.g., a smartphone) in charge of
the activity recognition; or ii) scenario composed of a high
performance device (e.g., a smartphone) that monitors and performs
light to moderate computing tasks (\ie~activity recognition) but
transmits the data to a back-end server, as a data center, for storage
or high-demanding computation. Many commercial smartphone applications
running activity recognition tasks compute these in the server
side~\cite{akimura2012compressed}---which makes this task independent
of the wearable device but, on the contrary, demands frequent wireless
data transmission. Offloading data from nodes to Cloud servers
remains into the Mobile Cloud Computing (MCC)~\cite{khan2013towards}
and Remote Health Monitoring (RHM) paradigms, leading to high energy
and economic savings~\cite{santambrogio2015power}. In this paper we
focus on the
\emph{scenario i} (one \emph{Plug\&Play} sensing node and a smartphone
as back-end). However, the methodology developed also fits in the
\emph{scenario ii}.

Developing energy-efficient system-level solutions and computationally
simple embedded software are warranted in order to drive the wearable
technology forward. At the hardware system level there are ingenious
solutions in the field of energy harvesting from kinetic energy using
piezoelectric~\cite{khalifa2017harke} and
capacitive~\cite{lan2017capsense} technologies that reduce
considerably the power consumption of the sensing nodes. They have
studied this for application in activity recognition problems, however
these solutions are still far from commercialization. In this paper,
we devise an ultra-low-power solution applicable to state-of-the-art
sensing nodes based on adaptive Compressed Sensing (CS) methodology to
reduce the amount of data being transmitted. Compressed sensing allows
us to create a representation of the original data in a transformed
domain reducing the amount of data to transmit with an acceptable
minimal information loss. Therefore, compressed sensing is more
applicable when having a periodic signal such as
ECG~\cite{mamaghanian2011compressed,chen2014compressed},
EEG~\cite{liu2015compressed}, PPG~\cite{baheti2009ultra} or motion
data.

Our work presents a low-power optimized temporal adaptive compressed
sensing framework for activity recognition. A preliminary version of
this study has been published previously
in~\cite{fallahzadeh2017adaptive}. To do this, a coarse-grained
activity recognition is performed on-node. By motion data, the system
i) implements coarse-grained node localization and ii) classifies
the \emph{signal type} to iii) automatically update the compressed
sensing rate in order to rigorously reduce the amount of data being
transmitted. We further implement our platform on a real device where
the energy consumption has been measured. Our results from real world
experiments show that the overhead of the methodology is negligible
compared to the overall energy saving achieved in wireless
transmission. Eventually, we compare the achieved savings to the
state-of-the-art compressed sensing approaches and the baseline.

The remainder of this paper is as follows. In~\secref{methodology}, a
general overview of the proposed methodology is described. Basic
concepts about algorithms and techniques applied are then introduce
in~\secref{preliminaries}. In~\secref{system}, the main modules of the
system are detailed.~\secref{results} shows the evaluation and the
discussion of the results. In \secref{discussion} the authors discuss
about the paradigm of real-time dependability in real-world
applications. Finally, conclusions of this work are drawn
in~\secref{conclusions}.



\section{Methodology overview}
\label{sec:methodology}
This section describes the high level overview of the proposed
system. Furthermore, we compare our framework to the traditional
compression techniques and the state-of-the-art techniques using
compressed sensing approaches.


An activity recognition framework consists of two differentiated
parts commonly described in most IoT systems: the sensing node and the
back-end computing unit. Our scheme consists of a resource constrained
and low cost activity tracking device and a higher performance
back-end, e.g., a smartphone. However, this methodology can also be
applied when the activity tracker is a smartphone and the back-end is
a server hosted in a data center. In the following, we compare
different compression strategies in sensing nodes, while the back-end
unit remains practically unmodified (as shown in~\figref{metho}).


\subsection{State-of-the-art methodologies}

The basic approach (\figref{base}) as opposed to compressed sensing,
uses conventional compression/decompression algorithms (such as ZIP
compression). Sensor reading are compressed locally, and are
decompressed and fed into a fine-grained activity recognition module
after transmission to a back-end station. The superiority of
compressed sensing over such conventional algorithms, in motion data
application, has been shown in literature~\cite{akimura2012compressed}
both in terms of computation complexity and compression performance.

The state-of-the-art approach (\figref{naive}) uses optimized
compressed sensing technique, but lacks any intelligent on-node
adaptivity. In other words, sensor readings are optimized using a
fixed compression rate (learned and optimized off-line). While it can
be conceived as a considerable improvement over the basic approach,
its minimization is limited because of the absence of sensitivity to
its context (\eg~varied attributes of the input data, on-body sensor
location). The performance of the framework proposed in this paper is
compared with the aforementioned framework in \figref{naive}, and also
with a baseline approach where data is transmitted without
compression.

In order to eliminate the real-time feedback dependency and improve
the energy efficiency of the activity recognition system, we propose a
temporal adaptive compressed sensing framework (\figref{metho}).

\ignore{
   The traditional sensing-compression approach is shown
   in~\figref{base}. In this approach the feedback---that is the label
   for user's current activity---is provided by the fine-grained
   activity recognition module after the signal is decompressed and
   reconstructed on the back-end server. This framework has real-time
   dependency on back-end connectivity and data transmission, which is
   a power consuming task, and in many cases---when connection between
   users' devices and the back-end server is not available---will not
   be applicable.  In this approach, data are acquired at Nyquist rate
   $n$, and $m < n$ data are sent after compression.

   \figref{naive} shows the version of the compressed sensing approach
   in the state-of-the-art. In this naive version, only the activity
   recognition is performed in the back-end. This approach however
   utilizes a fixed compression ratio for all inputs across all
   on-body locations and therefore lacks adaptivity.

}


\subsection{Proposed optimized methodology}
\label{sec:proposedMethod}

\begin{figure}
 \centering
 \subfloat[Basic compression scheme.]{
  \includegraphics[width=1.0\columnwidth]{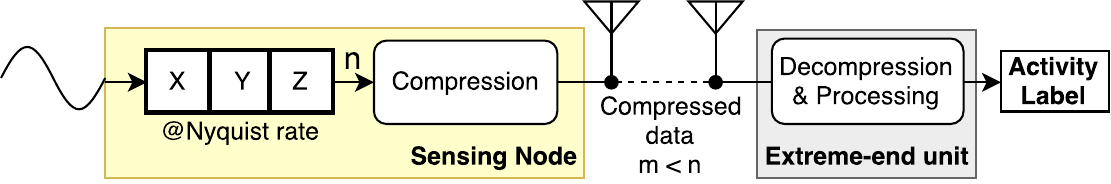}
  \label{fig:base}
 }
 \\
 \subfloat[Naive compressed sensing activity recognition system in the state-of-the-art.]{
  \includegraphics[width=1.0\columnwidth]{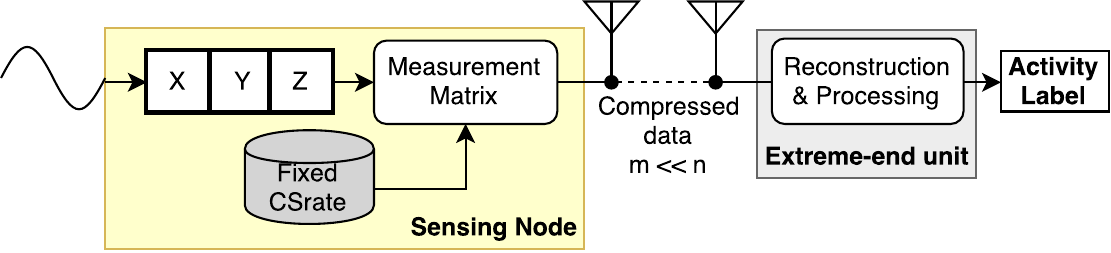}
  \label{fig:naive}
 }
 \\
 \subfloat[Adaptive temporal compressed sensing activity recognition system.]{
  \includegraphics[width=1.0\columnwidth]{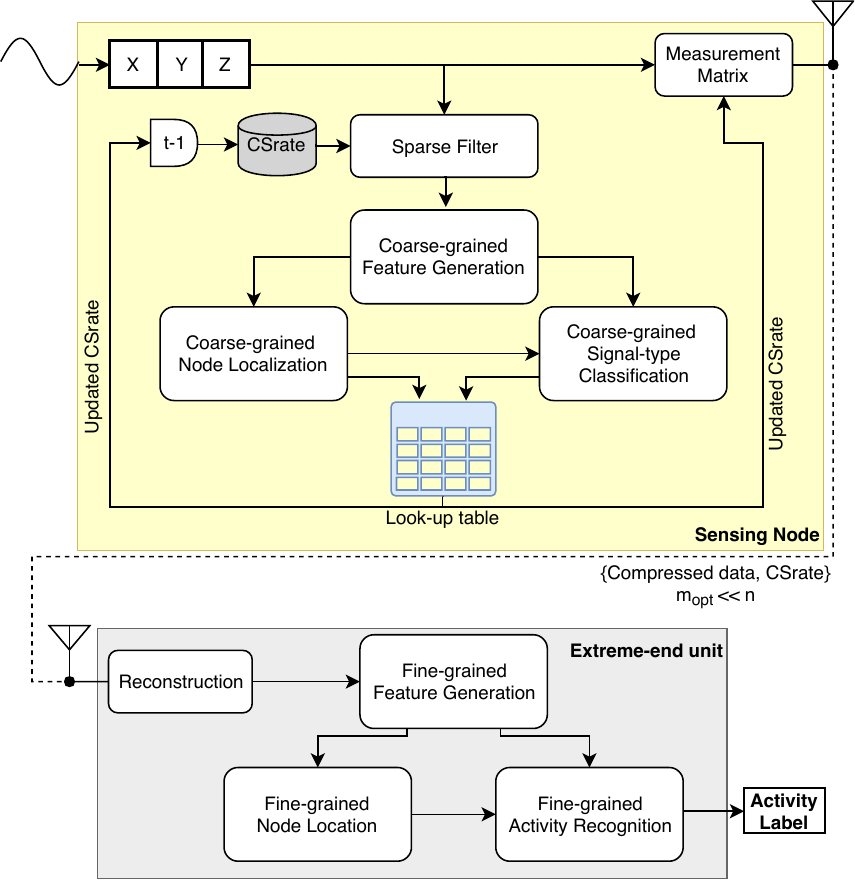}
  \label{fig:metho}
 }

 \caption{Overview of the basic, the state-of-the-art (naive) and the
 proposed compression methodologies. The back-end unit is detailed
 only in the last one.}
 
\label{fig:methods}
\end{figure}

The proposed adaptive compressed sensing activity recognition system
(\figref{metho}), inherits the capabilities of sate-of-the-art
approach and introduces novel low-cost coarse-grained input type
detection and on-body node localization algorithms. The
context-awareness provided by these two modules enables the node to
autonomously detect such changes and adjust the ratio and therefore
provide adaptive and robust optimal compression.

The sensing node can be potentially worn anywhere on the body: pants
pockets, chest, arm, and wrists are among the most common on-body node
location in activity recognition wearables (for simplicity throughout
the paper we refer to \emph{on-body node location}
as \emph{location}). In our case, the\ignore{It is required an
easy-to-use monitoring device, as users can turn the device on and
start to use it (i.e. a \emph{Plug\&Play} device). This} sensing node
performs a novel dynamic compressed sensing scheme to reduce the
amount of data sensed transmitted and thus enhance the battery
life. The computation server (\ie~back-end) receives the data and
reconstructs an approximation of the original signal to apply activity
recognition techniques after that on the application side.

Compressed sensing technique can reduce the amount of data being
sensed, processed, and transmitted with minimal impact on the
performance of the accuracy of the system (see~\subsec{cs}). This
technique can be applied at i) at sampling
rate~\cite{razzaque2014energy} (adaptive hardware sampling), or ii)
once the data are acquired at an adequate sampling rate.

The sensing node utilizes the sensor readings to extract features for
continuous context learning. Every time a data segment is gathered and
processed, the sensing node, if needed, re-configures the compression
ratio. Data segments are values of several seconds of a triaxial
accelerometer. Features will be extracted from the compressed
data. Using these features two tasks are performed: i) coarse-grained
node localization and ii) coarse-grained and location dependent signal
type detection as shown in~\figref{metho}.

We note that we make a distinction between the activity label and the
signal type. A activity label (\ie~physical activity label) is any
activity or movement that the subject wearing the device performs,
such as: walk or jump. On the other hand, a signal type is an abstract
and undetermined concept that refers to the idea that, signals having
different morphology and origin, can lead to an identical data
processing regardless of their corresponding activity labels. Thus,
many different activities can share the same signal type and therefore
the same compression ratio can be applied. From the perspective of a
sensing node worn on an arm, there is little to no difference between
the two activities with respect to compression ratio. Using the signal
type instead of simply relying on the activity label enables us to
leverage adaptivity in an unsupervised fashion, meaning we do not need
to label the training/test data to learn the signal type.

Under this circumstance, two optimization problems can be defined: i)
detecting the optimal number of signal types (clusters of signals per
on-body location), and ii) assigning a compression ratio to each
signal cluster such that the energy consumption is minimized. For each
location and signal type, a specific compression ratio will be applied
to the raw data according to the look-up table (that is trained
offline) stored in the local memory of the sensing node as shown
in~\figref{metho}. The input of this look-up table is the context
(\ie~signal type per location) learned by the course-grained module
and the output is the context optimized compression ratio. The
compressed data and the updated ratio are sent to the back-end unit,
where the compressed raw data are recovered. In the back-end, features
are extracted again to compute: i) fine-grained node localization, and
ii) fine-grained coarse-grained location-aware activity recognition.

In~\secref{preliminaries}, we briefly elaborate i) how the compressed
sensing technique is performed; ii) how to devise an offline GE-based
optimization algorithm; and finally iii) how to perform the activity
recognition task. All the modules in~\figref{metho} are further
elaborated in~\secref{system}.

\ignore{

Depending on the activity label we will require more or less
signal quality, or it might be that different activities need the
same quality level whatever it is. We talk then of kind of signal
instead of kind of activity and in this work we are going to figure
out how many kinds of signals we can distinguish, to apply
different compression ratios to each one.

Activity recognition algorithms are well known in the literature
(sports...). Extracting features from motion sensors we can apply
data mining techniques to classify and label different
movements. Motion sensors can be placed in different parts of the
body and still we will have to classify activity.

In the case of smartphones, worn location may vary as activities
do. So detecting the location and the kind of signal we will be able
to apply different compression levels to save energy.

We will be able to apply coarse-grained node location detection and coarse
signal-type detection from features extracted from compressed data.  }



\section{Preliminaries}
\label{sec:preliminaries}

\subsection{Compressed sensing}
\label{subsec:cs}
Compressed sensing theory states that a signal $x$ can be recovered
with minimal information loss by sub-sampling the signal with order $m
<< n$ where $n$ denotes the Nyquist sampling frequency under sparsity
and incoherence
assumption~\cite{candes2008introduction,eldar2012compressed}.

Compressed sensing---as many other compression techniques, such as
Wavelet or Fourier Transform---are efficiently applied in periodic
signals. Representation of periodic signals in terms of these
transformations lead to sparse signals; meaning that the most of the
coefficients $c$ will be equal to zero after the transformation. In
general, it can be asserted that the movement signals used in this
work can be considered periodic signals. We briefly describe how
compressed sensing is formulated and applied in our study.

Given $x$ sampled at Nyquist rate $n$, the signal $x$ can be
written as a linear combination as~\eref{signalx}:
\begin{equation} 
    x = \Psi c,
    \label{eqn:signalx}
\end{equation}

\noindent where $c$ is the set of sparse coefficients of the signal
$x$ after domain transformation given the transformation basis
$\Psi$. In our case, $x$ is a one dimensional sensor reading segment
of length $p$, $\Psi \in \mathbb{R}^{p \times p}$ is a Discrete Cosine
Transformation (DCT) matrix. Similarly, the length of $c$ is equal to
$p$.

To represent the sub-sampled signal $y$, the matrix $\Phi$ is
used. $\Phi$ is the \emph{sensing matrix} and takes $q$ random samples
of the signal $x$, thus:

\begin{equation} 
  y = \Phi x
  \label{eqn:signaly}
\end{equation}

In our case, signal $y$ is the data being sent from the monitoring
device to the back-end unit, and its length is $q \times 1$; and $\Phi
\in \mathbb{R}^{q \times p}$. Notice that $q/p = m/n$, and it is referred to
as the compression ratio $cr$. Randomization affects only mildly to
the calculus of features because these are based on statistics of the
morphology of the signal and not on temporal properties
(see~\subsec{data}).

To recover an approximation $x'$ of the original signal $x$ in the
back-end, the estimation of coefficients $c$ is
required. Using \esref{signalx}{signaly}, we can also write the
compressed signal $y$ as~\eref{signaly2}:

\begin{equation} 
  y = \Phi \Psi c = A c
  \label{eqn:signaly2}, 
\end{equation}

\noindent where $A \in \mathbb{R}^{q \times p}$ is the \emph{measurement
matrix}, defined in~\eref{measurementmatrix}. $A$ is the extraction of $q$
rows of the DCT matrix $\Psi$, and it is known by the sensing node and
the receiver-end to compute the estimation $c'$ of coefficients
$c$. Thus, the reconstructed signal $x' = \Psi c' \approx x$.

\begin{equation} 
  A = \Phi \Psi
  \label{eqn:measurementmatrix}
\end{equation}

Computation of coefficients of $c'$ is expensive and has to take place
in back-end, with higher performance than the monitoring
devices. According to the authors that drove the compressed sensing
techniques to its current
status~\cite{donoho2006compressed,candes2006robust}, each column of
matrix $A$ implies to solve a linear system. This is an undetermined
linear system and is hard to
solve~\cite{natarajan1995sparse}. Therefore, the problem needs to be
relaxed to $l_{1}-norm$ minimization:

\ignore{
\begin{equation} 
  min_{c'} \|c'\|~~s.t.~~\|y - Ac'\| < \epsilon,
  \label{eqn:l1}
\end{equation}
}

\begin{equation} 
   Minimize~~\|c'\|
   \label{eqn:l1}
\end{equation} 

\noindent Subject to

\begin{equation} 
  \|y - Ac'\| < \epsilon,
  \label{eqn:l1_1}
\end{equation}

\noindent where $\epsilon$ is the margin of reconstruction error. $c'$ can be found in polynomial time using linear programming.

In this work we propose a temporal adaptive compressed sensing
approach for activity recognition. To the best of our knowledge, only
one work by Yuan~\etal~\cite{yuan2013adaptive} presents a time
adaptive compressed sensing approach (applied to a video
problem). Chiu~\etal~\cite{chiu2015jice} pose adaption of basis
functions of the sensing matrices and adaption of the compression
ratio to the problem (energy auditing networks for different
services); however, the adaptation is not dynamic.

Compressed sensing is a very well known methodology in the field of
image processing. Hardware compressed sensing~\cite{duarte2008single}
has been commercialized recently. Unfortunately, there are no
hardware implementation in microcontrollers used in embedded wireless
monitoring devices, and only a RISC architecture of a
specific-application processor for compressed sampling is described by
Constantin~\etal~in~\cite{constantin2012ultra}. As a result, in this
paper, compressed sensing is implemented as firmware (FW) of the
monitoring node.


\subsection{Application in Activity Recognition}
\label{subsec:activity}

\begin{figure}
 \centering
 \subfloat[Three seconds of walking data.]{
  \includegraphics[width=1.0\columnwidth]{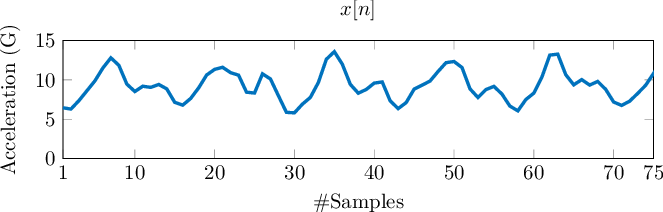}
  \label{fig:dctCompressedA}
 }
 \\
 \vspace{-0.3cm}
 \subfloat[Sparse DCT representation.]{
  \includegraphics[width=1.0\columnwidth]{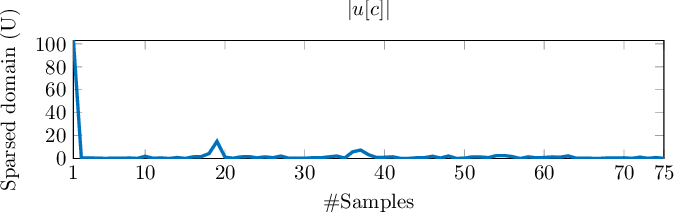}
  \label{fig:dctCompressedB}
 }
 \\
 \vspace{-0.3cm}
 \subfloat[Reconstructed signal.]{
  \includegraphics[width=1.0\columnwidth]{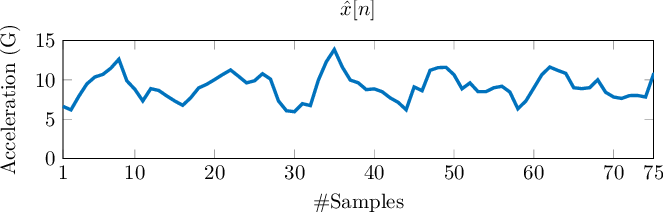}
  \label{fig:dctCompressedC}
 }
 \caption{Sparse recovery of activity \emph{walking} using $cr=33\%$ and $NRMSE=0.08$}
\label{fig:dctCompressed}
\vspace{-0.4cm}
\end{figure}

\figsref{dctCompressedA}{dctCompressedC} are examples of the processing 
of the compressed sensing algorithm.~\figref{dctCompressedA}
represents data from one axis of the accelerometer acquired at $25
Hz$. In~\figref{dctCompressedB} the sparse DCT is shown. This signal
is sparse enough to be transmitted---decreasing the amount of data and
resulting in significant power savings. Randomized sampling of 33\% of
this signal is used to recover $x'$ in~\figref{dctCompressedC}, as an
approximation of the original data $x$. It can be observed that the
morphology of the reconstruction is still distinguishable.

As Yang\etal~state in~\cite{yang2010energy}, the sparsity of the
physical activities varies for different types of movements and
locations. The authors in~\cite{yang2010energy}, evaluate different
compression ratios for different activities; they do not evaluate the
accuracy over the activity recognition but over the signal
reconstruction, and they conclude that this can lead to similar
information loss when measured by the Normalized Root-Mean-Square
Error (NRMSE). However a supervised classifier (trained with a large
enough labeled data) is required to detect changes in activities. We
hypothesize that optimization of the compression ratio based on
signal-type recognition may result in even more signification energy
savings while using an unsupervised approach.

\ignore{
~\figref{dctExamples} shows two movement randomly sampled and
compressed and reconstructed at ratio of 75\%. Despite the NRMSE are
quite high (0.16 and 0.25 in~\figref{dctExamples}a)
and~\figref{dctExamples}b) respectively), .
}

\ignore{

   \begin{figure}
    \centering \includegraphics[width=1.0\columnwidth]{dctExamples} \caption{\hl{DCT
    examples}} \label{fig:dctExamples}
   \end{figure}

}

\subsection{Optimization using Grammatical Evolution}
\label{subsec:GA}
As mentioned in~\secref{proposedMethod}, the system running on the
sensing node needs to detect the signal type---based on the detected
location---in order to apply the optimum \emph{cr}.

In our framework, it is not necessary to know the activity labels or
any other labeled data to work. It is important to highlight that we
do unsupervised classification (clustering) and it is one of the
strengths of the proposed model as opposed to the labor-intensive
supervised classification whose performance is bounded by the provided
training data. Instead, this model uses unlabeled data which is
abundant and easy to gather. First, we need to optimize our clustering and then we can find the minimum ratio
per each discovered cluster. The details of this bi-objective optimization
problem are given by the following problems:

\ignore{As there is no
direct relation between types of activities and types of signals and,
despite the different activities are known the number of types of
signals is unknown, the detection of the different types of signals
will be one of the goals to reach.
}

\begin{prob}[Optimal signal-type clustering per on-body location]
\label{prob1}
	For all the different activities from each on-body node
	location, find the clustering solution that optimally
	identifies the types of signals. See details
	in~\subsubsec{cgstc}.
\end{prob}
\begin{prob}[Optimal compression ratio per cluster]
\label{prob2}        
	For each cluster containing a particular signal type per
	on-body location, find the optimal compression ratio that
	minimizes the energy consumption of the node and maximizes the
	accuracy of the activity recognition in the receiver-end.
\end{prob}

In order to tackle these problems, we use
GE~\cite{ryan1998grammatical}. GE is a grammar-based form of Genetic
Programming (GP), used to generate programs in any language. GE's
internal behavior is based on a Genetic Algorithm (GA). GAs are stochastic methods to solve complex optimization
problems. Its heuristic search allows to tackle faster big
combinational problems, as ours, the optimal solution or really close
to this. GE has genotypes---represented as integer numbers---that
select production rules from a group expressed in a Backus Naur Form
(BNF) and lead to phenotypes. A phenotype is a tree-shape structure
which is evaluated in an iterative process. In our case, a phenotype
represents the set of activity-features to be selected.

\begin{figure}[t]
 \centering
 \subfloat[Example of population. Mix and mutation of individuals are shown.]{
  \includegraphics[width=1.0\columnwidth]{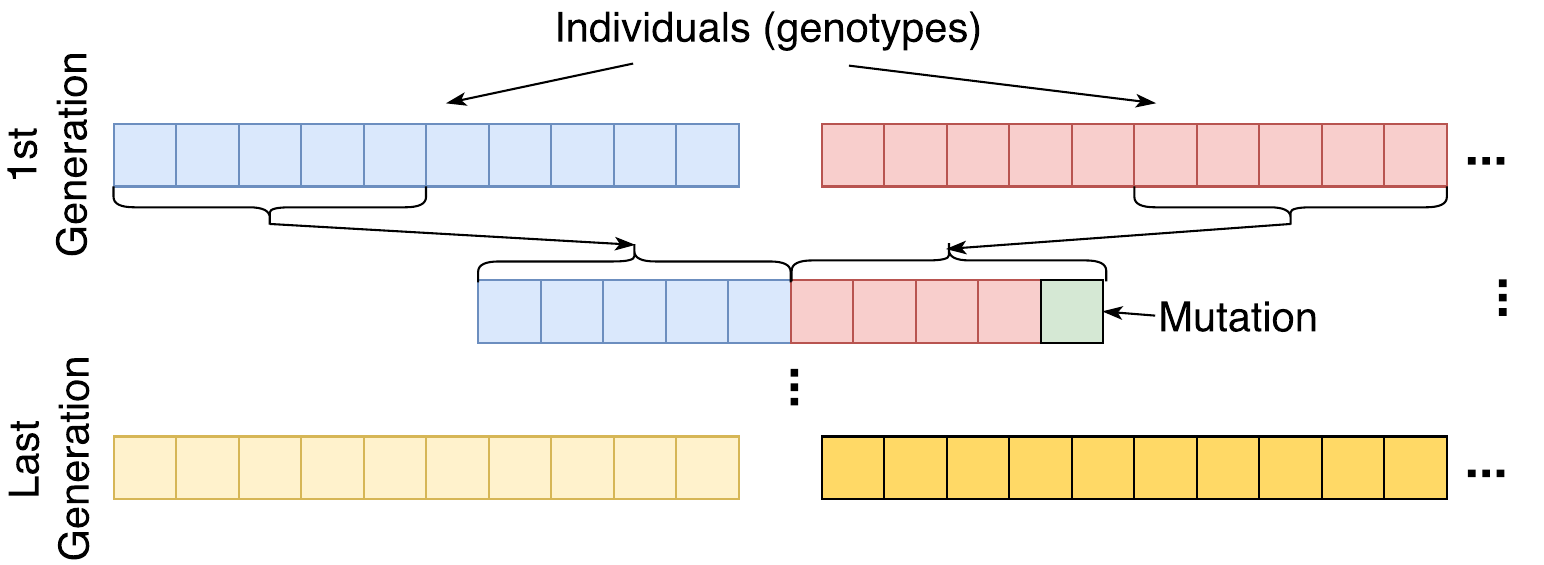}
  \label{fig:population}
 }
 \\
\vspace{-0.4cm}
 \subfloat[Chromosome array and its decoded final expression.]{
  \includegraphics[width=1.0\columnwidth]{decodingGE}
  \label{fig:decodif}
 }
 \caption{Example of the population and chromosome decodification of the GE problem.}
\label{fig:ge}
\vspace{-0.3cm}
\end{figure}

\begin{figure}[]
 \centering
 \includegraphics[width=0.9\columnwidth]{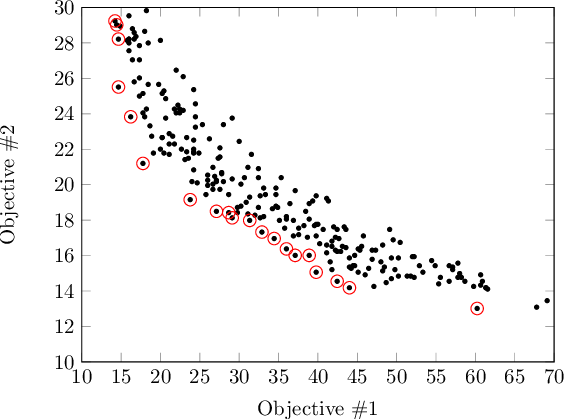}
 \caption{Example of Pareto front to minimize a two-dimensional
 objective problem.}
\label{fig:pareto}
\vspace{-0.4cm}
\end{figure}

Following with the biological simile, a GA evolves a population formed
by a set of individuals (the genotypes) as shown
in~\figref{population}. Individuals mutate and mix with each other to
create new ones in every generation. Each individual is evaluated, and
those of each generation that better fit the objective will survive
and evolve with a higher probability in future generations. For an
example of a complete process of decodification of a genotype refer
to~\cite{pagan2016grammatical}. The gray array in the top
of~\figref{decodif} shows how a genotype looks like. After applying
the BNF's rules, in our case, a phenotype looks like the white array
shown in the bottom of~\figref{decodif}. The designed grammar is shown
in BNF format in~\figref{BNF}. This grammar allows the optimization
of the number of clusters for the signal type detection.

\begin{figure}[]
\begin{lstlisting}[basicstyle=\scriptsize\ttfamily,breaklines=true,frame=tb]
N = { <Model>, <K>, <Boolean> }
T = { TRUE, FALSE, 2, 3,..., 25 }
S = <Model>

P = {
  I    <Model>     ::= [<K>, <Boolean>, <Boolean>,...,
                       <Boolean>] # Till the total number of features F
  II   <K>         ::= 2|3|4...|25
  III  <Boolean>   ::= TRUE|FALSE
}
\end{lstlisting}
\caption{BNF grammar used for to solve the problem of optimal signal
  type classication.}
\label{fig:BNF}
\vspace{-0.4cm}
\end{figure}


A BNF grammar is represented by a set of parameters in the form $\{N,
T, P, S\}$, where $N$ denotes the set of non-terminals (coded
symbols), $T$ denotes the set of terminals (decoded expressions), $P$
denotes the set of production rules to substitute the elements of $N$
into $T$, and $S$ is a non-terminal element of $N$ used as starting
symbol. A model (labeled as \texttt{Model}) is an array combination of
(terminals, $T$): the number of clusters (\texttt{K}), and a set
of \texttt{F} Boolean values (\texttt{Boolean}) that indicate which
activity-features $\in \texttt{F}$ have to be selected.

Because our approach is based on a multi-objective optimization, we
use a multi-objective genetic algorithm inside GE: the Non-dominated
Sorting Genetic Algorithm II (NSGA-II)~\cite{deb2002fast}. The main
difference between a simple GA and NSGA-II is that at the end, as
opposed to a single best solution, a set of non-dominated solutions is
obtained (traditionally named as Pareto front). NSGA-II is an elitist
approach, what means that a small part of the best candidates remain
unchanged into the next generation---they remain as parents of the
next generation, which enhance convergence and allows computation of
an approximation of the entire Pareto front. \figref{pareto}
illustrates an example of a two-dimensional Pareto front. Black points
in~\figref{pareto} are the last individuals or solutions of the
NSGA-II execution. Red circles surround those solutions that lead to
the Pareto front. Refer to~\cite{colmenar2011multi} for details on
multi-objective GE implementation.

\ignore{
NSGA-II is an elitist approach, what means that a small part of the
best candidates remain unchanged into the next generation---they
remain as parents of the next generation. Each individual of the
population plots a vector---3D and 2D vectors in both of the
optimization problems respectively. The NSGA-II
algorithm~\cite{deb2002fast} will choose the best ones that are in the
non-dominated Pareto front. An individual belongs to the Pareto front
and is called non-dominated if any of its objectives can be improved
(minimized in our case) without degrading the others. \figref{pareto}
represents a two-dimensional Pareto front. Black points
in~\figref{pareto} are the last individuals or solutions of the GA
problem (the best ones). Red circles surround those solutions that
lead to the Pareto front.
}


\subsection{Optimizing the clustering scheme}
\label{subsec:cluseval}

As mentioned in~\subsec{GA}, the optimization problem $\#1$ finds the
optimum number of clusters that matches the different number of signal
types. To perform this task, a \emph{k-means++} clustering is utilized
(see~\subsubsec{cgstc}). \emph{k-means++} is an unsupervised approach
that needs an internal evaluation index to rank each individual
(solution) of the GE.

In order to evaluate each individual solution of the GE, an internal
evaluation index is needed. There are several internal indices for
evaluating a cluster, such as: the Davies-Bouldin index, Dunn index or
silhouette coefficient. While any of these indexes could be used in
our optimization, the former has been employed because of its lower
complexity. The Davies-Bouldin index~\cite{davies1979cluster} is given
by~\eref{daviesbouldin}:

\begin{equation}
  \overline{R} \equiv \frac{1}{K}\sum_{i = 1}^{K} \max_{i \ne j}\left(\frac{\sigma_{i} + \sigma_{j}}{d(c_{i}, c_{j})}\right)
  \label{eqn:daviesbouldin}
\end{equation}

\noindent where the parameter $K$ denotes the total number of clusters, the parameter $c_{i}$ denotes the centroid of
cluster $i$, the parameter $\sigma_{i}$ represents the average
distance of all elements in cluster $i$ to the centroid $c_{i}$, and
the parameter $d(c_{i}, c_{j})$ is the distance between centroids
$c_{i}$ and $c_{j}$. We use the Euclidean distance as the distance
metric.

The parameter $\overline{R}$ is non-negative. Lower values of
$\overline{R}$ indicate better clustering. $\overline{R}$ relates the
scatter of the cluster with the distance between clusters. As the
distance between clusters is in the denominator
in~\eref{daviesbouldin}, $\overline{R}$ is low when clusters are not
spread and each one is compact (\ie~the distance between clusters
increases).



\section{System detail}
\label{sec:system}
In this section, we elaborate each module in~\figref{metho} from the
perspective of the two functional components of the system: the node,
and the back-end computing unit.


The optimization problems have been solved using an implementation of
GE and NSGA-II algorithms available in the HERO Java
Library~\cite{HERO}. Compressed sensing techniques, \emph{k-means++}
clustering, and clustering evaluation have been computed in
Matlab. The decompression has been carried out
through the method to recover sparse signals via convex programming
implemented in Matlab by Romberg and developed in~\cite{candes2005l1}.
The activity recognition and the node localization have been computed
using WEKA~\cite{hall2009weka}.

\subsection{Sensing node view}
\label{nodeview}
Sensing nodes acquire data from triaxial accelerometers and store them
into segments ($s$) of few-second sensor data. Nodes can be placed on
different locations of the body. For the sake of simplicity, five
known locations are considered. For the compressed sensing approach,
each data segment $s$---at time $k$---is sub-sampled at rate
$cr[t-1]$, leading to a new segment $s'$. This takes place in
the \emph{Sparse Filter} module (see~\figref{metho}). From this
segment, different features are extracted in the \emph{Coarse-grained
Feature Generation} block. With these features, the location is
determined in the \emph{Coarse-grained node Localization} module. For
each location the signal type is detected by the \emph{Coarse-grained
Signal Type} module. With the location and the signal type, the
compression ratio is updated to $cr[t]$ according to the
embedded \emph{Look-up table}. The measurement matrix $A$
from~\eref{measurementmatrix} is applied in the \emph{DCT} module and
compressed data are transmitted.

\subsubsection{Sparse Filter}
In this module, the sub-sampled $s'$ sequence is created picking
$cr[t-1]$ random samples of segment $s$. At this point, the
compression ratio has not been updated yet. Therefore, transitory
states that lead to the selection of non-optimal and delayed
compression ratios, may occur between transition of
activities. However, these transitions are sparse in time and are
insignificant in practice.

\subsubsection{Feature generation}
Using the sub-sampled sequence $s'$, 30 morphological features are
computed, as it will be shown in~\subsec{data}. Computation of these
features is computationally light and a relatively simple task to
perform for resource constrained monitoring nodes.

\subsubsection{Coarse-grained node localization}
 
Different data mining algorithms have been trained to classify and
detect the node localization: classification trees, support vector
machines or \emph{k-means} among others. \emph{Random
Forest}~\cite{breiman2001random} has shown the highest accuracy using
10 cross-fold validation. \emph{Random Forest} is a supervised
algorithm that generates a set of different trees under certain
rules. To classify a new instance---vector of features extracted from
sequence $s'$---the generated trees vote for the most popular
class. All classes (locations) must be in the training set.

\emph{Random Forest} algorithm is a sequence of \emph{if-else} statements
which makes it more practical to implement in sensing nodes with
constrained computation capacity. The \emph{Random Forest} algorithm
is trained offline using all 30 features extracted from uncompressed
raw data. In real time, as the features are extracted from sub-sampled
data, a coarse-grained node localization will take place.


\subsubsection{Coarse-grained signal type classification}
\label{subsubsec:cgstc} 

Several known daily and sports activities are considered for this
research (see~\subsec{data}). As defined in~\secref{proposedMethod},
we will classify our movement data into signal types.  As was
previously mentioned, the cardinality of the set of signal types is
unknown (signal type detection is an unsupervised classification
problem). To achieve the best performance-power saving trade-off, we
need to solve the optimization problems~\ref{prob1} and~\ref{prob2}
stated in~\subsec{GA}:

{\bf Optimization problem \#1:}
Clustering of the optimum number of signal types maximizing the
number of clusters $\#k~\in~K$, the quality $Q_{idx}$ of the final
cluster, and minimizing the number of required features $\#f~\in~F$.

As granularity criterion, we consider $K=2,3,...,25$, and $\#k>1$ as a
constraint. This granularity has been established in order to minimize
the computation time of the optimization problem. The limit is defined
by the upper-bound on the number of clusters possible: 125
($\Delta_{min}cr=100*\frac{1}{125}=0.8\%$), equal to the size of input
data. Maximizing the number of clusters will allow us to apply several
compression ratios that lead to a more accurate control of the energy
consumption in the sensing node. However, high granularity could lead
to close clusters with slightly different compression ratios. To avoid
this, we have established a trade-off between the number of clusters
and granularity, and thus, we studied only 25 different values, so
that the minimum increment of
$\Delta_{min}cr=100*\frac{25}{125}=4.0\%$, which is not too low or not
too high (for 25 Hz and 5 seconds data length it leads to compression
of 5 samples). The quality of the cluster is computed using the
Davies-Bouldin index, thus $Q_{idx}~=~\overline{R}$. $\#f$ will be
minimized to reduce the computation in the node, with
$F=1,2,...,30$. Given that, the multi-objective problem can be
formulated as:

\begin{equation}
  \min\left(\#f, Q_{idx}, \frac{1}{\#k}\right)
  \label{eqn:optcluster}
\end{equation}

Given a number of clusters $\#k \in K$ and a set of features $f' \in
F$, the system computes the goodness of the clustering
$Q_{idx}$. Because there are no labels to indicate the different
signal types, this is an unsupervised clustering that has been
implemented using a \emph{k-means++} algorithm.

\begin{figure}
 \centering \includegraphics[width=1.0\columnwidth]{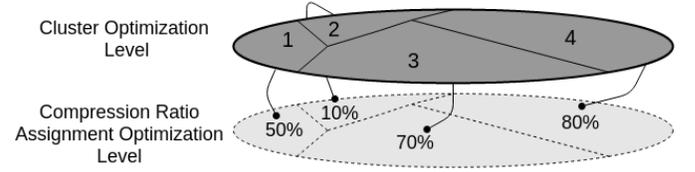} \caption{Per
 each location: division of the different signal types found and the
 assignment of the compression ratios.}  \label{fig:signalTypeCR}
\end{figure}

\figref{signalTypeCR} illustrates two optimization levels. 
The upper one is an example of the optimal partitioning of the data
space. In this example, four different types of clusters and the
corresponding values are found.

In this optimization, the selected solutions are named as
$S_{P_{ST}}^{*}$. For each one of the selected solutions a compression
ratio set is assigned. These appear in the lower abstract plane
in~\figref{signalTypeCR}, and these values are found through a second
optimization problem stated in~\subsec{GA}.

{\bf Optimization problem \#2:}
To find the optimum set of compression ratios $cr_{set,f'}$ for a
given set of features $f'$ that maximizes the weighted mean of the
compression ratio $\overline{cr}_{set,f'}$ and maximizes the accuracy
of the activity recognition $\alpha_{AR}$ in the back-end. Thus,
the multi-objective problem states that:

\begin{equation}
\min\left(1-\overline{cr}_{set}, 1-\alpha_{AR}\right)
\label{eqn:optcluster2}
\end{equation}

The weighted mean compression ratio is given by:

\begin{equation}
\overline{cr}_{set,f'} = \frac{1}{N}\sum_{i=1}^{\#k} n_{i} cr_{i,f'},
\label{eqn:cr}
\end{equation}

\noindent where $N$ is the number of instances in the training set, $n_{i}$
is the number of instances that have been classified into the cluster
$i$, and $cr_{i,f'}$ is the compression ratio for cluster $i$ using
the set of features $f'$. This optimization---based on each solution
$S_{P_{ST}}^{*}$---leads to a new Pareto front. In this Pareto front, the
selected solutions $S_{P_{CS}}^{*}$ must satisfy:

\begin{equation}
\max_{S_{P_{CS}}^{*}}~(\overline{cr}_{set,f'}~|~\varepsilon_{AR} \le \varepsilon_{th})
\label{eqn:finalSols}
\end{equation}

\noindent to be implemented in the nodes. 

\eref{finalSols} establishes that the solutions to be implemented 
are the ones with higher weighted mean compression ratio from those
solutions which have an error in the activity recognition
$\varepsilon_{AR}$ less than a threshold $\varepsilon_{th}$. The
threshold $\varepsilon_{th}$ is a quality criterion set defined as the
extra error added to the lower bound error in activity recognition
$\varepsilon_{AR} = 100-\alpha_{AR_{Base.}}$. Here, the extra error
has been set to $5\%$. Thus, $\varepsilon_{th} = \varepsilon_{AR} +
5\%$. This value has been chosen based on criterion and experience of
the researchers. It has been considered that this value helps to relax
conditions to achieve better consumption metrics, as well as to keep a
considerable quality level.

The maximum number of clusters established in the previous
optimization problem also sets the granularity in the compression
ratios. Thus, in the $cr_{set,f'}$ each value $cr_{i} \in
CS=0,4,8,...,96$---$K=25$ equidistant levels in the interval [0, 100).

As in the example, incoming data classified as signal type 1
in~\figref{signalTypeCR}, is compressed with $cr=50\%$, while data
classified as signal type 4 is compressed with $cr=80\%$.

These two optimization problems are computed offline, and the
relations \emph{\{localization, signal type, cr\}} are stored in the
internal memory of the node in a \emph{Look-up table}.

\subsubsection{Measurement matrix}
When the output of the previous module leads to a new $cs[t]$
compression ratio, this is used to select the
corresponding \emph{measurement matrix}. Raw data are multiplied by
the DCT matrix and then sent wirelessly to the back-end unit.


\subsection{Back-end view}
\subsubsection{Fine-grained feature generation}
A segment of compressed data that includes information of the
\emph{cr} used is received. With this information, the
estimation $x'$ of the original signal is computed using a copy of the
measurement matrix $A$ stored on the node. Features are computed from
the recovered segment to perform the node localization and the
activity recognition.

\subsubsection{Fine-grained node localization}
In this phase we perform node localization on the received and
reconstructed data segments. The \emph{Random Forest} procedure has
been chosen as in the sensing node. Once the data are recovered, the
system works independently of the compression ratio used in the
sensing node. The \emph{Random Forest} is trained using the data
recovered for all compression ratios. This leads to a fine-grained
node localization. All the 30 features are considered in
the \emph{Random Forest} algorithm.

\subsubsection{Fine-grained activity recognition}
\label{subsubsec:actRecog}

As in~\subsubsec{cgstc}, after training different algorithms, such as
multi-layer perceptrons or random trees, a \emph{Random Forest}
algorithm has shown the best performance in terms of training
error. After the node localization phase, the fine-grained activity
recognition is performed. All features are considered for this purpose
in the back-end.



\section{Evaluation and results}
\label{sec:results}
In this section, we thoroughly evaluate the proposed system. We
present and discuss the results and the performance of the two
optimization problems: i) clustering of the optimum number of types of
signals; and ii) optimum assignment of compression ratios for those
clusters in addition to the modules in the sensing nodes and the
back-end units. Finally, a realistic experimental framework is
proposed and used to measure the energy savings of the system.

\subsection{Data}
\label{subsec:data}
To evaluate the framework, the real world human activity
dataset \emph{UCI Daily and Sports Activities Data
Set}~\cite{barshan2014recognizing} is used. These are real data
acquired from 8 subjects (4 male and 4 female, between the ages 20 and
30) asked to perform 19 daily and sports activities, such as: sitting,
walking, running at different speed, jumping, rowing or playing
basketball. Data are acquired at $25 Hz$ using a triaxial
accelerometer. Each activity takes 5 minutes and data are chopped into
non-overlapping segments of 5 seconds (125 samples per axis). Data
were acquired simultaneously from a wearable sensor placed on five
different on-body locations: torso (T), right arm (RA), left arm (LA),
right leg (RL) and left leg (LL). The data set was split into training
($80\%$) and test ($20\%$).

From these data, 10 different statistical features were
extracted~\cite{ghasemzadeh2015power} for each axis of the
accelerometer. This leads to 30 features per data segment of 5
seconds. These features are: the amplitude of the segment $amp$, the
median $med$, the mean $mn$, the maximum and minimum values ($max$ and
$min$), the peak to peak increment $p2p$, the variance and the
standard deviation ($var$ and $std$), the root mean square value $rms$
and, finally, the difference between the first and last values of the
sequence $s2e$. In~\subsec{features} our feature selection mechanism
is described. Furthermore, the impact in energy of this selection is
elaborated in~\subsec{energy}.


\subsection{Optimization}
\label{subsubsec:optresults}
In this section, the results of the optimization problems using GE and
NSGA-II are shown. For both processes, the number of individuals for
each generation has been set to 250. Likewise, the number of
generations has been set to 1000 and 500 for signal-type clustering
optimization and compression ratio assignment optimization,
respectively. For each generation, all individuals are evaluated
externally using compiled code created using Matlab. The number of
individuals per generation has been chosen to reduce the amount of
external calls to the classifiers. The number of generations has been
chosen after some tests. For these values of the parameters, the
solutions remain approximately stable and no new solutions
(individuals) appear at the end of the optimization process.


\subsection{Performance}
\label{subsubsec:performance}

Following paragraphs show the performance results of the optimization
problems on-node and the \emph{Random Forest} classifier for both node
localization and activity recognition in the sensing node and the
back-end unit.

\subsubsection{Coarse-grained feature generation}

This module computes 10 features for each axis of the
accelerometer. Features are calculated over compressed data which
allows important reduction of the energy consumption.

\subsubsection{Signal-type clustering optimization}

\begin{figure}[]
 \centering
 \subfloat[Torso.]{
  \includegraphics[width=0.6\columnwidth]{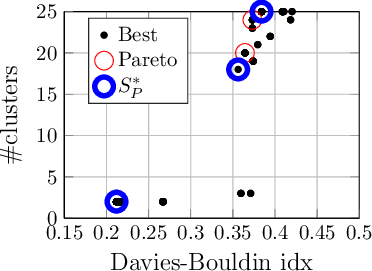}
  \label{fig:optClustT}
 }
 \\
 \subfloat[Right arm.]{
  \includegraphics[width=0.6\columnwidth]{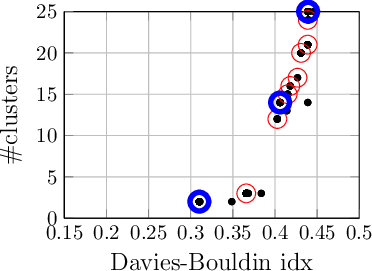}
  \label{fig:optClustRA}
 }
\\
 \subfloat[Left arm.]{
  \includegraphics[width=0.6\columnwidth]{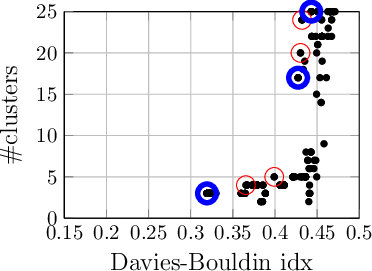}
  \label{fig:optClustLA}
 }
 \\
 \subfloat[Right leg.]{
  \includegraphics[width=0.6\columnwidth]{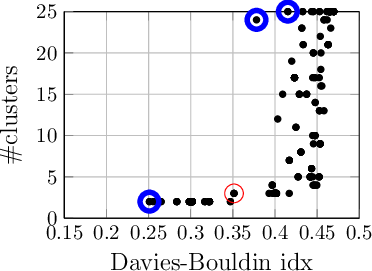}
  \label{fig:optClustRL}
 }
 \\
 \subfloat[Left leg.]{
  \includegraphics[width=0.6\columnwidth]{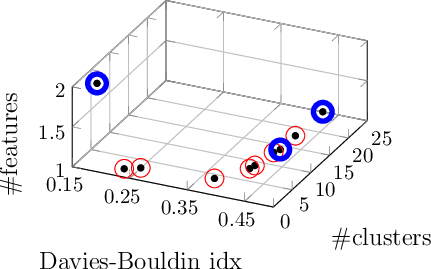}
  \label{fig:optClustLL}
 }
 \caption{Signal-type clustering optimization.}
 \label{fig:Clustopt}
\end{figure}

Solutions that solve~\eref{optcluster} are plotted
in~\figtoref{optClustT}{optClustLL}. These figures represent the best
solutions (black points) of the GA,~\ie~the last generation of the
population. The non-dominated solutions $S_{P}$ represent the Pareto
front (circles).

Because we have three optimization goals, the figures should be
plotted in three dimensions. However, since all the solutions on the
Pareto front share the same value on feature axis $\rho_{\#f}=1$, we
plotted the figure in 2D. The only exception is the
location \emph{Left leg} that uses two features. All the 10 solutions
in~\figref{optClustLL} belong to the non-dominated Pareto
front. Reducing the number of features is a significant result because
it diminishes the complexity of the implementation of
the \emph{k-means} algorithm in the sensing node, and the same time
maximizes the performance of our resulting system.

It is worth mentioning that the minimization of the number of features
was easily accomplished by the NSGA-II algorithm---from the first
generations the selection was rapidly established. This can be
understood as another indicator that not all the features are useful
and only a small selection of them becomes meaningful. In a additional
mono-objective optimization process it was observed that the optimal
number of clusters for our training set is $\#k_{opt}=2$. As our
criterion to reduce the energy costs goes through increasing this
number, the optimization process achieves solutions using a high
number of clusters appeared in the Pareto front (in the next section
it is shown that the criterion to maximize $\#k$ is correct).

In the Pareto fronts we have selected two extreme solutions and a
third one between them to compute the next optimization process. These
selected solutions $S_{P_{ST}}^{*}$ are represented as blue bold
circles in~\figtoref{optClustT}{optClustLL}. The Davies-Bouldin index
$\overline{R}$ remains in the same range for all locations but for
\emph{torso}, where it is lower. For all the observations in the heuristic
experiments, the values $0.15 \le \overline{R} \le 1.60$; so results
are considered acceptable being always lower than $0.5$, and lower
than $0.45$ for the selected solutions $S_{P_{ST}}^{*}$.

\tblref{finalClusters} shows the number of unique compression ratio values
$\#cr_{unique}$ for each on-body location. We define the relative
clustering-ratio assignment efficiency ($\nu_{cluster}$) for each
location as the ratio of the number of unique compression ratio
assignments over the number of distinct clusters, as follows:

\begin{equation}
   \nu_{cluster}=100\frac{\#cr_{unique}}{\#k_{merged}} (\%)
\label{eqn:crUnique}
\end{equation}
    
\begin{table}[h!]
\caption{Values of optimized objectives per location.}
  \centering
\begin{tabular}{ccc|ccc}
\hline
Location & $f'$ & $\#k$ & $\#k_{merged}$ & $\#cr_{unique}$ & $\nu_{cluster} (\%)$ \\ \hline
T & \emph{mnX} & 18 & 17 & 10 & 58.8 \\ 
RA & \emph{mnX} & 14 & 13 & 8 & 61.5\\ 
LA & \emph{medY} & 25 & 24 & 13 & 54.2\\ 
RL & \emph{mnZ} & 2 & 2 & 2 & 100\\ 
LL & \emph{mnZ} & 14 & 14 & 11 & 78.6\\ 
\hline
\end{tabular}
\label{tbl:finalClusters}
\end{table}

Smaller values of $\nu_{cluster}$ indicate more redundancy, meaning
that the same compression ratios appear several times while clusters
cannot be merged in \figref{signalTypeCR}, because the signal types
for that location are quite similar. On the contrary, larger
$\nu_{cluster}$ means that a the identified clusters are more varied
in terms of the compressed sensing ratios assigned to them, and the
look-up table in \figref{metho} is smaller as well. As an
example, \tblref{centroids} lists the distinct clusters for Torso
location and their associated compression ratios. According to the
obtained efficiency, we can observe that the granularity criterion
established for the number of clusters $\#k$ was computationally
sufficient.
  
\begin{table*}[htbp]
\caption{Clusters for Torso.}
\centering
\resizebox{2.0\columnwidth}{!}{
   \begin{tabular}{cccccccccccccccccc}
   \hline
   Cluster & A & B & C & D & E & F & G & H & I & J & K & L & M & N & O & P & Q \\ \hline
   \emph{cr (\%)} & $72$ & $64$ & $72$ & $12$ & $72$ & $44$ & $76$ & $40$ & $68$ & $72$ & $56$ & $64$ & $56$ & $76$ & $88$ & $64$ & $28$ \\
   Centroid (units) & $-6.74$ & $-4.67$ & $-3.72$ & $-3.14$ & $-1.84$ & $-0.59$ & $0.36$ & $1.43$ & $2.51$ & $4.23$ & $6.99$ & $7.69$ & $8.38$ & $8.96$ & $9.25$ & $9.51$ & $9.75$ \\ \hline
   \end{tabular}
}
\label{tbl:centroids}
\end{table*}

\subsubsection{Compression ratio assignment optimization}
\label{subsubsec:cr}

\begin{figure}[]
 \centering
 \subfloat[Torso.]{
  \includegraphics[width=0.6\columnwidth]{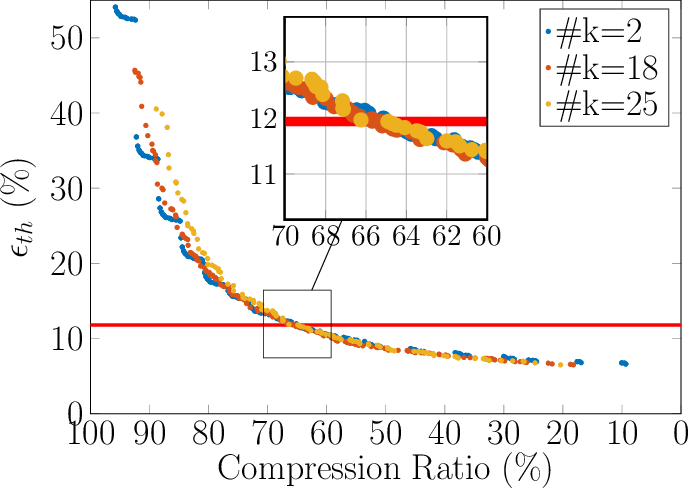}
  \label{fig:optCSratioT}
 }
 \\
 \subfloat[Right arm.]{
  \includegraphics[width=0.6\columnwidth]{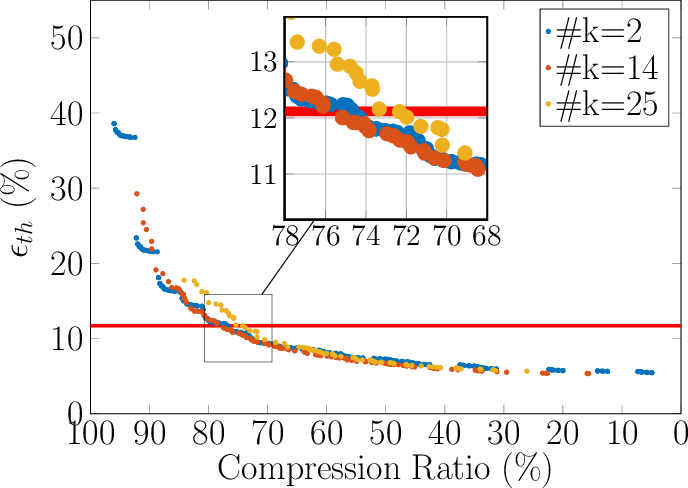}
  \label{fig:optCSratioRA}
 }
\\
 \subfloat[Left arm.]{
  \includegraphics[width=0.6\columnwidth]{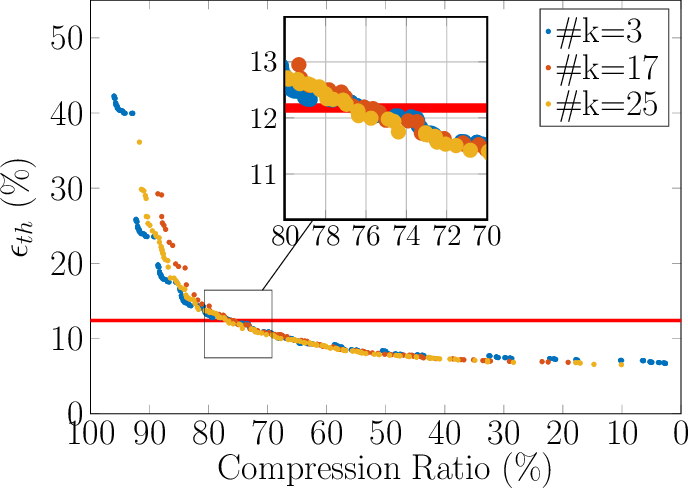}
  \label{fig:optCSratioLA}
 }
 \\
 \subfloat[Right leg.]{
  \includegraphics[width=0.6\columnwidth]{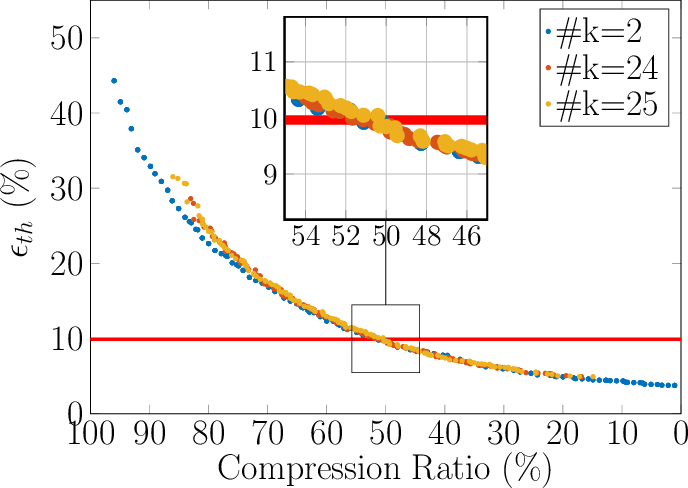}
  \label{fig:optCSratioRL}
 }
 \\
 \subfloat[Left leg.]{
  \includegraphics[width=0.6\columnwidth]{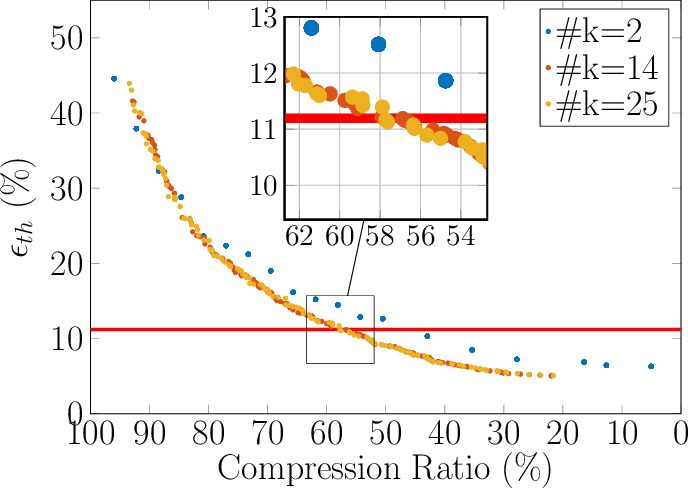}
  \label{fig:optCSratioLL}
 }
 \caption{Compression ratio assignment optimization.}
 \label{fig:Clustopt}
\end{figure}

For each solution $S_{P_{ST}}^{*}$, another NSGA-II optimization has
been performed. The Pareto front resulting from each optimization over
the training set is shown in~\figtoref{optCSratioT}{optCSratioLL}.

\begin{table*}[t!]
\caption{Accuracy of the selected solutions after the two optimization processes and the naive approach compared to the baseline.}
  \centering \resizebox{2.0\columnwidth}{!}{
\begin{tabular}{c|ccc|ccc|ccc|ccc|ccc||c}
\hline
Location & \multicolumn{3}{c|}{T} & \multicolumn{ 3}{c|}{RA} & \multicolumn{ 3}{c|}{LA} & \multicolumn{ 3}{c|}{RL} & \multicolumn{ 3}{c||}{LL} & Average $\alpha (\%)$\\ \hline

$\alpha_{AR_{Base.}} (\%)$ / $\varepsilon_{th} (\%)$ & \multicolumn{ 3}{c|}{93.2 / 11.8} & \multicolumn{ 3}{c|}{93.3 / 11.7} & \multicolumn{ 3}{c|}{92.6 / 12.4} & \multicolumn{ 3}{c|}{95.1 / 9.9} & \multicolumn{ 3}{c||}{95.0 / 10.0} & 93.8\\ \hline

\#k & 2 & \textbf{18} & 25 & 2 & \textbf{14} & 25 & 3 & 17 & \textbf{25} & \textbf{2} & 24 & 25 & 2 & \textbf{14}  & 25 & \multirow{3}{*}{89.0}\\ 

$\overline{cr}_{set,f'} (\%)$ & 64.9 & \textbf{65.2} & 64.6 & 76.7 & \textbf{77.4} & 75.4 & 76.0 & 75.7 & \textbf{76.6} & \textbf{51.2} & 50.6 & 50.3 & 37.5 & \textbf{42.8} & 40.1 \\ 

$\alpha_{AR_{Adap.Temp.}}$ $| \overline{cr}_{set,f'} (\%)$ & 88.2 & \textbf{88.4} & 88.3 & 88.6 & \textbf{88.6} & 88.2 & 87.7 & 87.6 & \textbf{87.7} & \textbf{90.2} & 90.2 & 90.4 & 90.2 & \textbf{90.0} & 90.0 \\ \hline

\multicolumn{ 1}{r|}{$\alpha_{AR_{Naive}} (\%)$ $| cr= \textbf{42.8} \%$} & \multicolumn{ 3}{c|}{91.9} & \multicolumn{ 3}{c|}{91.5} & \multicolumn{ 3}{c|}{92.1} & \multicolumn{ 3}{c|}{89.9} & \multicolumn{ 3}{c||}{89.6} &  \textbf{91.0}\\ 

\multicolumn{ 1}{r|}{$| cr=62.6 \%$} & \multicolumn{ 3}{c|}{89.0} & \multicolumn{ 3}{c|}{89.8} & \multicolumn{ 3}{c|}{90.1} & \multicolumn{ 3}{c|}{83.6} & \multicolumn{ 3}{c||}{83.5} & 87.2\\ 

\multicolumn{ 1}{r|}{$| cr=77.4 \%$} & \multicolumn{ 3}{c|}{83.6} & \multicolumn{ 3}{c|}{86.7} & \multicolumn{ 3}{c|}{87.7} & \multicolumn{ 3}{c|}{76.1} & \multicolumn{ 3}{c||}{76.7} & 82.2\\ 

\hline

\end{tabular}
}
\label{tbl:solsCS}
\end{table*}

For the criterion stated in~\eref{finalSols}, horizontal lines have
been drawn in~\figtoref{optCSratioT}{optCSratioLL}. These lines cross
through the three Pareto fronts and the results are compared
in~\tblref{solsCS} for each location. This table compares the accuracy
of the baseline approach, our adaptive temporal compressed sensing
methodology and three different naive solutions---the minimum, maximum
and average value of the selected solution (bold numbers) from all
locations. Adjusting $\varepsilon_{th}$ changes the optimal
compression ratios, respective to the derivative of the Pareto
front. In \figtoref{optCSratioT}{optCSratioLL}, it can be seen that, a
larger value for $\varepsilon_{th}$ leads to more compression and
energy saving, and lower number of clusters. On the contrary, smaller
$\varepsilon_{th}$ leads to less compression and energy saving, and a
higher number of clusters ($\#k$).

As expected, in most of the cases, a high number of clusters leads to
high weighted mean compression ratios, and thus, a lower energy
consumption. The compression ratios are different for each node
location and similar for few locations---arms and legs---because
movements are similar. The highest compression ratios are achieved for
nodes located on arms. Therefore, it is expected that nodes placed on
arms save more energy due to data transmission.

We note that the same compression ratio could be assigned to different
clusters.  Those with the same compression ratio that are adjacent are
merged in $\#k_{merged}$, and the new centroid is computed. After this
process, $\#k$ has been reduced at least in one for three locations
(see~\tblref{finalClusters}).


The highest, the lowest and, the average compression ratios of the
solutions selected for each location have been studied in the naive
methodology. The last three rows in~\tblref{solsCS} compare their
accuracy. Only the solution with $cr= 42.8\%$ is compliant with the
quality criterion stated in~\eref{finalSols} ($5\%$ threshold on
residual error of classification). This is the solution selected to
compare with our proposed methodology. So, henceforth, the energy
consumption due to transmission for the naive case and LL must
coincide.

\subsubsection{Coarse-grained node localization}

For the node localization process performed in the sensing node,
a \emph{Random Forest} algorithm was trained. As memory is a limited
resource in monitoring devices, only one \emph{Random Forest} model is
stored. This model was trained with data without compression from all
locations. The location is detected when compressed data are processed
by this module. The accuracy of the trained model reaches up to
$97.8\%$ as shown in the confusion matrix
in~\tblref{coarsedLocationConfMat}.

\begin{table}[]
  \caption{Confusion matrix of the coarse-grained node localization module~($\%$)*.}  
   \begin{tabular}{rccccc}
    \cline{2-6}
    Classified as $\rightarrow$ & T & RA & LA & RL & LL \\ \cline{2-6}
    T & 97.8 & \cellcolor{gray}0.3 & \cellcolor{gray}1.6 & \cellcolor{blue!8}0.0 & \cellcolor{blue!8}0.3\\
    RA & \cellcolor{blue!8}2.7 & 91.8 & \cellcolor{blue!8}4.3 & \cellcolor{blue!8}1.2 & \cellcolor{blue!8}0.0 \\
    LA & \cellcolor{blue!8}4.5 & \cellcolor{gray}1.2 & 91.2 & \cellcolor{blue!8}1.5 & \cellcolor{blue!8}1.6 \\
    RL & \cellcolor{gray}0.0 & \cellcolor{gray}4.8 & \cellcolor{gray}0.1 & 83.3 & \cellcolor{blue!8}11.8 \\
    LL & \cellcolor{gray}0.9 & \cellcolor{gray}1.5 & \cellcolor{gray}7.1 & \cellcolor{gray}14.3 & 76.1 \\     \cline{2-6} \\
  
  \end{tabular}
  \label{tbl:coarsedLocationConfMat}
  \hspace{0.7cm} \footnotesize{*Light gray means sub-compression. Dark gray means over-compression.}
\end{table}

Mislocalizations happen mostly between legs. The higher error happens
when $14.4\%$ of times LL is classified as RL and $11.8\%$ of times RL
is classified as LL. This will lead to over-compression and
sub-compression respectively---according to the selected solutions
in~\tblref{solsCS}. These dissimilarities might compensate each other
regarding the accuracy and the energy consumption. Therefore,
according to our results in~\tblref{coarsedLocationConfMat}, we show
that this module does not propagate the error because of compensation.

\subsubsection{Fine-grained node localization}

When data is reconstructed in the back-end, features are computed
and node localization is performed. The accuracy of the fine-grained
node localization module is shown in the second column
in~\tblref{fineLocation}, and they are the best classification of the
node location possible in the back-end if there were no
misclassifications in the coarse-grained node localization.
These results are compared with the ones achieved using the naive 
system with a fixed compression ratio of $cr= 42.8 \%$ and the
baseline approach as well.

A \emph{Random Forest} algorithm has been trained for each approach
and the results are shown in~\tblref{fineLocation}. As can be seen,
results are really close each other and there are only slight
differences with respect to the baseline. Our approach differs only
$1\%$ accuracy from the naive methodology---despite the average
compression ratio of our approach is $62.6\%$ compared with $42.8\%$
of the naive method.

\begin{table}[ht!]
\caption{Accuracy of node localization in back-end unit $(\%)$.}
   \resizebox{1.0\columnwidth}{!}{
  \centering \begin{tabular}{cccc}
\hline
\multirow{2}{*}{If node placed in} & Temporal adaptive & Naive & \multirow{2}{*}{Baseline} \\ 
& compressed sensing & $cr=42.8 \%$ &  \\ \hline
Torso & 97.4 & 98.2 & 98.6 \\ 
Right arm & 97.6 & 98.7 & 98.7 \\ 
Left arm & 95.4 & 97.1 & 97.5 \\ 
Right leg & 95.1 & 95.6 & 97.7 \\
Left leg & 93.2 & 93.8 & 96.7 \\ \hline
Average & 95.7 & 96.7 & 98.2 \\\hline
\end{tabular}
}
\label{tbl:fineLocation}
\end{table}


\subsubsection{Fine-grained activity recognition}
The actual results for the activity recognition are those shown
in~\tblref{solsCS}. $\alpha_{AR_{Baseline}}$ represents the baseline
accuracy for the activity recognition algorithms trained in the
back-end of the system. As there are no limitations in the
computation, in the back-end all features are computed and
used. $\alpha_{AR_{Adap.Temp.}}|\overline{cr}_{set,f'}$ represents the
final accuracy for the activity recognition for our methodology. The
final solutions are those in bold in~\tblref{solsCS}. The average
$\alpha_{AR_{Adap.Temp.}}|\overline{cr}_{set,f'}$ for all locations s
$89.0\%$. As aforementioned, the accuracy of the naive solution
selected is $91.0\%$---$2\%$ higher that the average value of our
methodology. In spite of this, in~\subsec{energy} we show that, being
both methodologies compliant with the quality criteria, the benefits
of our proposal in terms of energy savings are significant.


\subsection{Energy consumption}
\label{subsec:energy}

To test the energy performance of the proposed methodology, an
experimental set-up has been developed. In this section we show the
results in terms of energy savings due to the data transmission (which
represent the $81\%$ of energy consumption in the baseline mode), and
we compare them with the small overhead that the system introduces for
the computation of the on-node calculations performed in our
custom-designed activity monitoring device shown in~\figref{metho}.


\subsubsection{Experimental set-up}
\begin{figure}
 \centering
 \includegraphics[width=0.6\columnwidth]{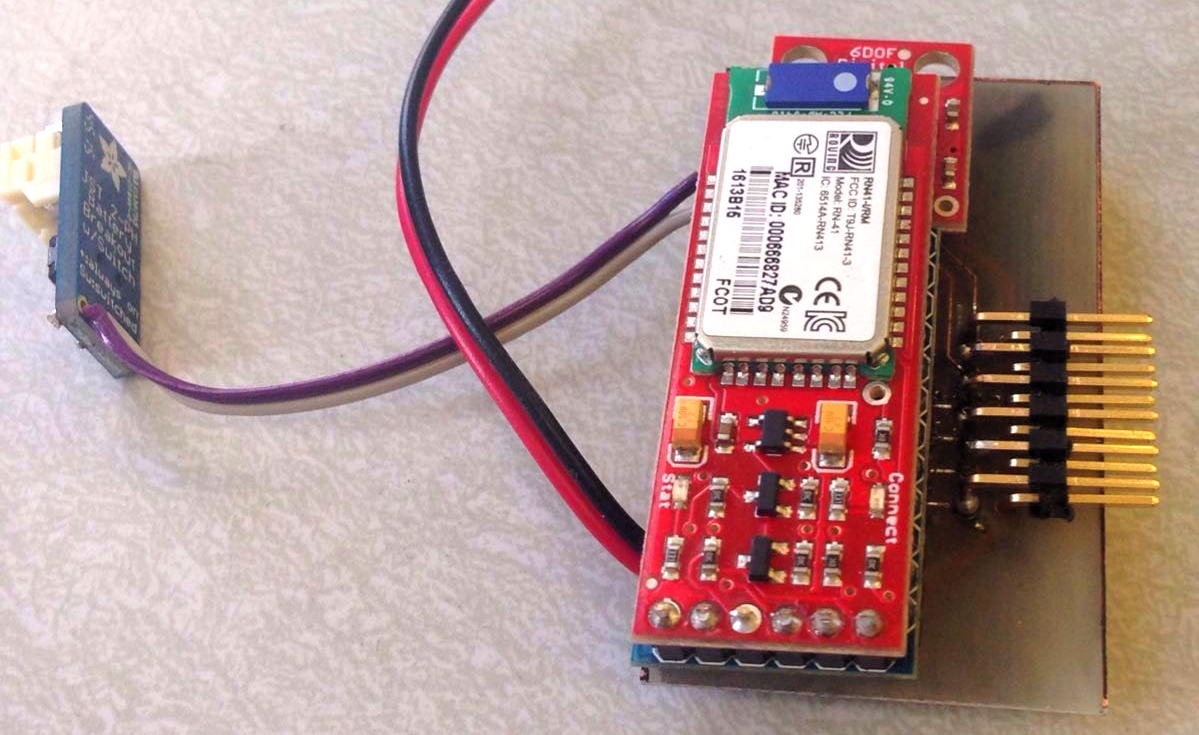}
 \caption{Experimental monitoring node.}
 \label{fig:node}
\end{figure}

In this study, we programmed our methodology onto an actual activity
  monitoring device (see~\figref{node}). The experimental set-up
  consists of a sensing device that performs the system’s tasks and
  another device to measure its power consumption. The sensing node
  incorporates one ATmega328 microcontroller running at $8$ MHz, a
  $10$-bit precision accelerometer sensor ADXL335, and a Bluetooth
  module RN41-3 that sends the data to a computer configured in deep
  sleep mode and using a power transmission of $-12dBm$. The device
  under measure is a platform that monitors the current in the sensing
  node's battery using the INA219 sensor. Current is measured through
  a shunt resistor of $0.1\Omega$ the measurement precision is $100\mu
  A$ and sent to the computer at a rate of $500Hz$. All results are
  computed for the weighted average compression ratio achieved for
  each location in~\tblref{finalClusters}.

To obtain the total energy consumption $\varepsilon_{T}$, a linear
energy model is proposed in~\eref{energy} and the results are compared
with the baseline energy consumption $\varepsilon_{Baseline}$, where
the dummy system only transmits raw data every 5 seconds. This model
considers the single energy consumption of each module of the system
(node side in~\figref{metho}) in three main levels: sensing $\sigma$,
processing $\pi$ and transmission $\tau$. We assume that the total
energy consumption is the sum of individual consumption of each module
of the system.

Tasks are executed independently in a loop. The execution time to
process 5 seconds of gathered data is also measured using
time-stamps. The drawn current is measured as well. Using the timing
information we are able to compute the energy consumption.

\begin{equation} 
     \varepsilon_{Total} = \sigma + \pi + \tau,
     \label{eqn:energy}
\end{equation}

\noindent with:

\begin{equation} 
     \pi = SF_{\varepsilon} + FG_{\varepsilon} + (NL_{\varepsilon} + ST_{\varepsilon}) + MM_{\varepsilon}
    \label{eqn:energyProcessing}
\end{equation}

In the naive approach

\begin{equation} 
  \pi_{Naive} = MM_{\varepsilon},
    \label{eqn:energyProcessing2}
\end{equation}

\noindent because no feature computation, on-body node localization,
or signal-type detection is performed.

To ensure that the system passes through all the states, a piece of
the test dataset has been coded in program memory. This data is read
once and stored in RAM as for real gathered data. To lead to the
energy model in~\eref{energy}, we proceed as follows:

\begin{enumerate}
\item To measure the consumption of the sensing process $\sigma$, the
accelerometer module is connected and a simple code reads movements at
$25 Hz$---same sampling rate than the UCI Data Set uses---and stores
data in RAM.

\item
$\pi$ is the result of the five different blocks in the node side. For
each location we individually measure the energy consumption of the
processing modules. As shown in~\eref{energyProcessing}: the
$SF_{\varepsilon}$ consumption of the \emph{Sparse Filter}, the
$FG_{\varepsilon}$ due to the \emph{Coarse-grained Feature
Generation}, both machine learning algorithms
blocks \emph{Coarse-grained Node Localization} ($NL_{\varepsilon}$)
and \emph{Coarse-grained Signal-Type} detection ($ST_{\varepsilon}$),
and the \emph{Measurements Matrix} using the DCT ($MM_{\varepsilon}$).

\begin{itemize}
\item
$SF_{\varepsilon}$: For each location is stored in ROM memory. It is a
1D array of $\overline{cr}_{set,f'}$ random numbers. Each number
represents an index of one element of 5 seconds of data stored at $25
Hz$. Its energy consumption is $SF_{\varepsilon}$.

\item
$FG_{\varepsilon}$: the consumption of this module depends on the
compression ratio, the larger the segment, the heavier the
computation. The consumption of this module will be calculated when
computing all 30 features.

\item
$NL_{\varepsilon}$: a \emph{Random Forest} is implemented. This is a
sequence of $if-else$ sentences. For the implementation using
$\#f=30$, the average size (number of nodes) of the trees is 2020. The
number of trees in the forest is 100. For the sake of simplicity we
coded one, but the total energy consumption takes it into account.


\item
$ST_{\varepsilon}$: this consumption is the result of a \emph{k-means}
algorithm. As the optimal number of features found for all locations
is $\#f=1$, this implementation finds the closer centroid within a 1D
group of $\#k_{merged}$ clusters in a sequence of $if-else$ sentences
(see~\tblref{finalClusters}).

\item
$MM_{\varepsilon}$: the multiplications of DCT matrices lead this
consumption. The measurement matrices are stored in ROM and loaded to
RAM at the beginning of the execution in order to perform the
multiplications faster.
\end{itemize}

\item
Finally, the model includes the energy consumption of data
transmission using Bluetooth, $\tau$. For these experiments:

\begin{itemize}
\item
First, we measure the baseline energy consumption
$\varepsilon_{Baseline}$ when our system is not implemented and all
raw data is transmitted every 5 seconds. Every second the Bluetooth
transmitter wakes up and sends 375 samples of data---25 samples for
each axis of the accelerometer using 10 bit precision. Thus, the
energy overhead of this baseline solution comes mainly from the
switching of the transmitter.

\item
The amount of samples to send every 5 seconds for each location
according to the weighted average compression ratios are: i) 132 for
T, ii) 87 for RA, iii) 90 for LA, iv) 183 for RL and v) 216 for
LL. Using the naive methodology 216 samples are sent every 5 seconds.
\end{itemize}
\end{enumerate}

\begin{table}[]
\caption{Average energy consumption and total energy savings.}
  \centering
  \resizebox{1.0\columnwidth}{!}{ 
\begin{tabular}{ccccccc}
\hline
Location & $\sigma (mJ)$ & $\pi (mJ)$ & $\tau (mJ)$ & $\varepsilon_{Total}$ & $\varepsilon_{Tot.Savings} (\%)$\\ \hline
T & 13.4 & 4.7  & 15.9 & 34.0 & 51.6 \\ 
RA & 13.4 & 3.5  & 11.2 & 28.1 & 60.0 \\                     
LA & 13.4 & 3.6  & 10.7 & 27.7 & 60.6\\                      
RL & 13.4 & 5.7  & 23.5 & 42.6 & 39.4\\                      
LL & 13.4 & 6.8  & 27.8 & 48.0 & 31.7\\ \hline               
Naive & 13.4 & 1.4 & 27.8 & 42.6 & 39.4\\ \hline         
Baseline & 13.4  & - & 56.9 & 70.3 & - \\ \hline                     
\end{tabular}
}
\label{tbl:energy}
\end{table}

\begin{figure}
 \centering
 \includegraphics[width=0.8\columnwidth]{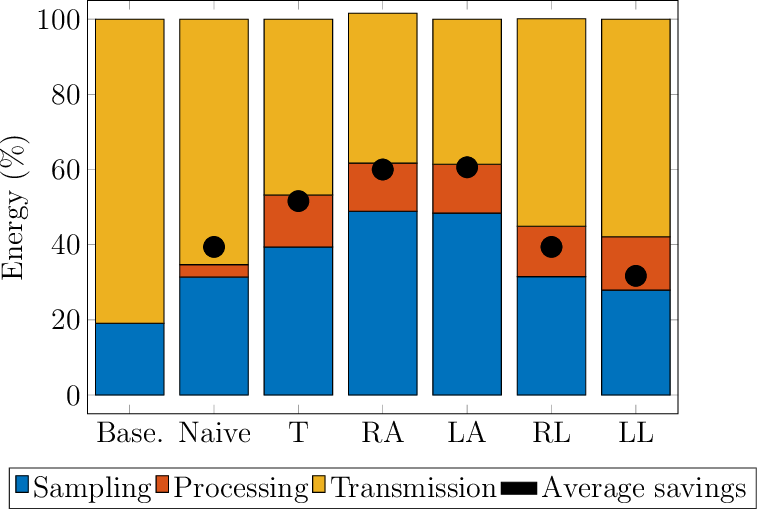}
 \caption{Relative energy consumption in the sensing node.}
 \label{fig:energy}
\end{figure}

With this experimental set-up, we obtain the energy consumption listed
in~\tblref{energy}. This table shows the average energy values per
second. The consumption of the sensing process $\sigma$ is an offset
equal for all locations.  $\pi$ is the average consumption of the
processing in the sensing node (see~\eref{energyProcessing}
and~\eref{energyProcessing2}). $\tau$ is the consumption due to data
transmission that takes the lower values for locations with higher
compression ratio.

As expected, the larger the weighted average compression ratio, the
higher the energy savings. This can be seen especially when node is
located in LA.~\figref{energy} shows the relative percentages of the
energy consumption for the tree components assumed in~\eref{energy},
for each location, the naive approach and the baseline. The energy
consumption of the most expensive process, the transmission, has been
reduced up to $81.2\%$ for LA, and a small overhead of our system
leads to total energy savings up to $60.6\%$ in this location when
compared with the baseline. Compared with the naive approach, for the
LA location too, these values have been reduced up to $61.5\%$ in
transmission, and $35.0\%$ for the overall system. All these solutions
and the naive one are compliant with the restriction of extra error in
activity recognition of $5\%$ over the baseline.

While the energy of computation might seem expensive, is a small
$\sim13\%$ overhead for all locations. Computation time ranges from $1.9$
to $2.2$ seconds, from lower to higher $\overline{cr}_{set,f'}$. All
consumption that performs the expression of $\pi$ are detailed
in~\tblref{energyPi}.
\ignore{Data transmission does not take long, but the
switching of the transmitter is expensive.}

\begin{table*}[]
\caption{Average processing energy consumption per second $\pi$ for $\#f=30$ and $\#f=6$ features.}
  \centering
\begin{tabular}{cccccccc|ccc}
\hline
\multirow{ 2}{*}{Location} & \multirow{ 2}{*}{$SF_{\varepsilon} (\mu J)$} & \multicolumn{2}{c}{$FG_{\varepsilon} (mJ)$} & \multicolumn{2}{c}{$NL_{\varepsilon} (mJ)$} & \multirow{ 2}{*}{$ST_{\varepsilon}(\mu J)$} & \multirow{ 2}{*}{$MM_{\varepsilon} (mJ)$} & \multirow{ 2}{*}{$\pi_{\#f=30} (mJ)$} & \multirow{ 2}{*}{$\pi_{\#f=6} (mJ)$}  & \multirow{ 2}{*}{$ \varepsilon_{\pi_{\frac{6}{30}Savings}} (\%)$}  \\ \cline{3-6}
& & $f_{30}$ & $f_{6}$& $f_{30}$ & $f_{6}$  &  &  \\ \hline

T & 4.0 & 3.4 & 2.9 & 0.5 & 0.6 & 0.2 & 0.8 & 4.7 & 4.3 & 8.5  \\ 
RA & 2.8 & 2.4 & 2.0 &  0.5 & 0.6 & 0.2 & 0.6 & 3.5 & 3.2 & 8.6 \\
LA & 2.9  & 2.5 & 2.2 &  0.5 & 0.6 & 0.2 & 0.6 & 3.6 & 3.4 & 5.6 \\
RL & 5.9 & 4.1 & 3.2 &  0.5 & 0.6 & 0.1 & 1.1 & 5.7 & 4.9 & 14.0 \\
LL & 6.9 & 4.9 & 3.8 &  0.5 & 0.6 & 0.1 & 1.4 & 6.8 & 5.8 & 14.7 \\ \hline
Naive & 6.9 & - & - & - & - & - & 1.4 & - & - & - \\ \hline

\end{tabular}
\label{tbl:energyPi}
\end{table*}

Computation of features is the most expensive process. Computation of
the \emph{Sparse Filter} ($SF_{\varepsilon}$) and \emph{Signal-type
clustering optimization} ($ST_{\varepsilon}$) are almost negligible.

Our analysis estimated computation power overhead of $13\%$ (of the
total energy consumed, for all locations). This is quite insignificant
as at the end, total energy savings of at least, $31.7\%$ was
achieved. The average saving over all locations in our approach was
$48.7\%$ which is considerably higher that the naive model
($39.4\%$). The only case where naive model outperformed our approach
was LL location. The reason behind this observation is that when
optimizing the compression ratio for the naive approach, the maximum
ratio belonged to LL and hence, for only that on-body location, the
naive approach can perform well without the need for an on-node
localization module. However, it is obvious that for any other
location this saving would degrade due to lack of on-node
location-awareness and intrinsic bias of the optimization model
towards the LL location. Our result is significant, because we were
able to achieve higher savings on average despite having computation
overheads (\eg~by having a coarse-grained node localization module)
while making the system much more robust against on-body sensor
displacement.
    
We have demonstrated that significant reduction of data transmission
of our adaptive temporal compressed sensing approach lead to energy
savings, resulting into substantial increments of battery life. We
have also demonstrated that the adaptive temporal version is better,
on average, than the naive approach compressing data at a fixed rate.

\subsection{Study of features}
\label{subsec:features}
\begin{figure}[t]
 \centering
 \includegraphics[width=0.8\columnwidth]{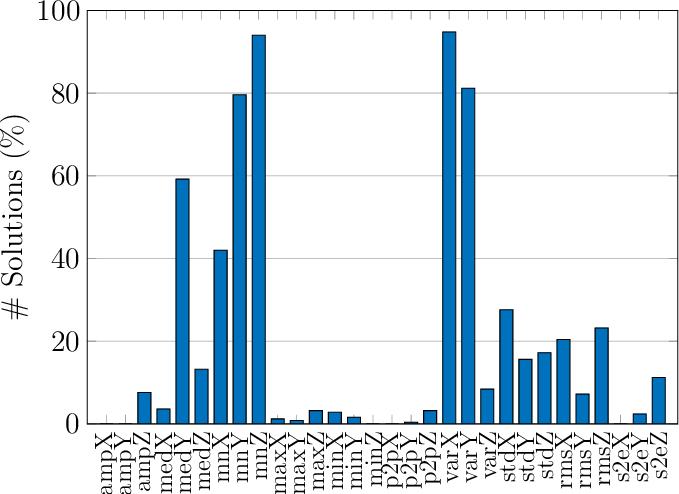}
 \caption{Histogram of features selected for the best GE solutions in the Optimization problem \#1.}
 \label{fig:histFeatures}
\end{figure}

As mentioned previously, $30$ features have been extracted. It was
noticed that some features were selected more times than others by the
optimization algorithms. An example is shown in the histogram shown
in~\figref{histFeatures}. This figure depicts the distribution of
features selected by the Optimization problem \#1 for both dominated
and not dominated solutions.

\begin{figure*}[ht]
 \centering
 \subfloat[Torso.]{
  \includegraphics[width=0.37\columnwidth]{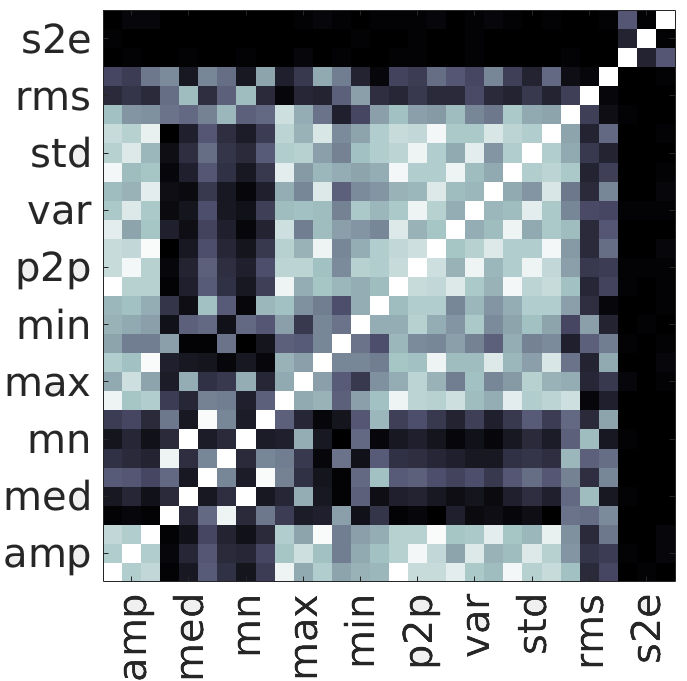}
  \label{fig:corrsT}
 }
 \subfloat[Right arm.]{
  \includegraphics[width=0.37\columnwidth]{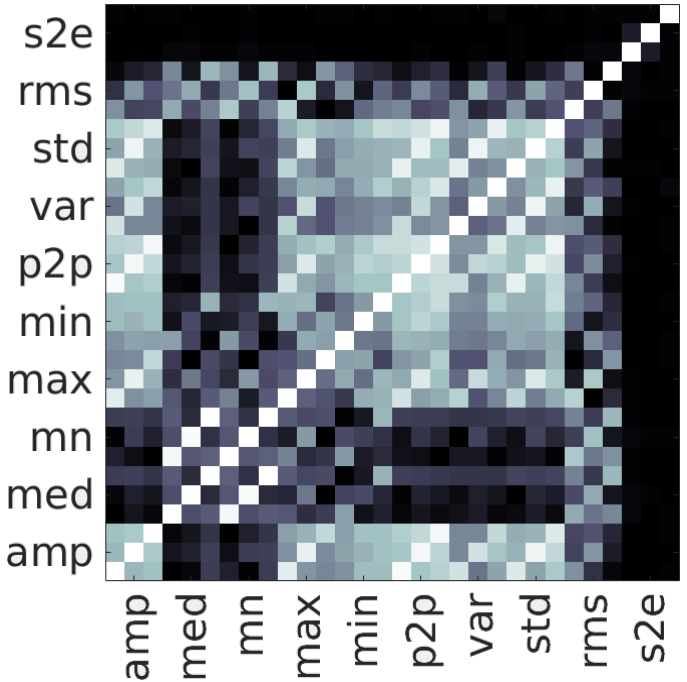}
  \label{fig:corrsRA}
 }
 \subfloat[Left arm.]{
  \includegraphics[width=0.37\columnwidth]{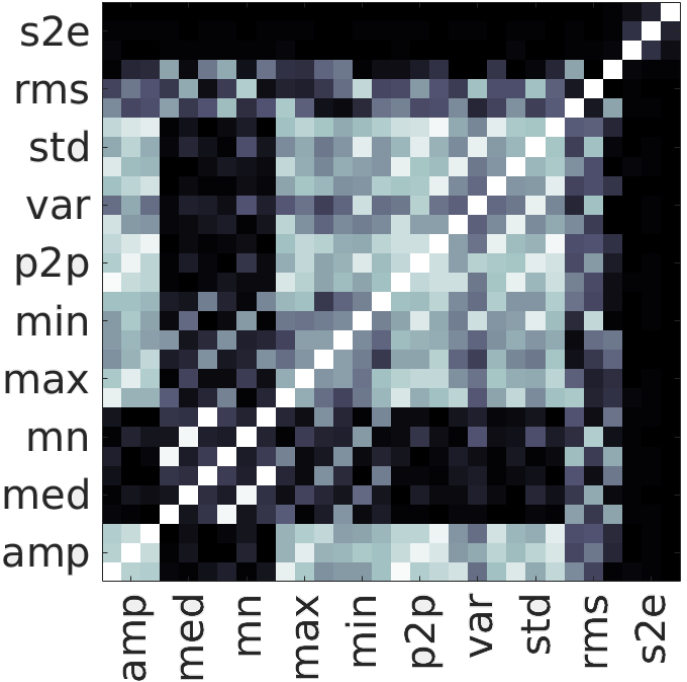}
  \label{fig:corrsLA}
 }
 \subfloat[Right leg.]{
  \includegraphics[width=0.37\columnwidth]{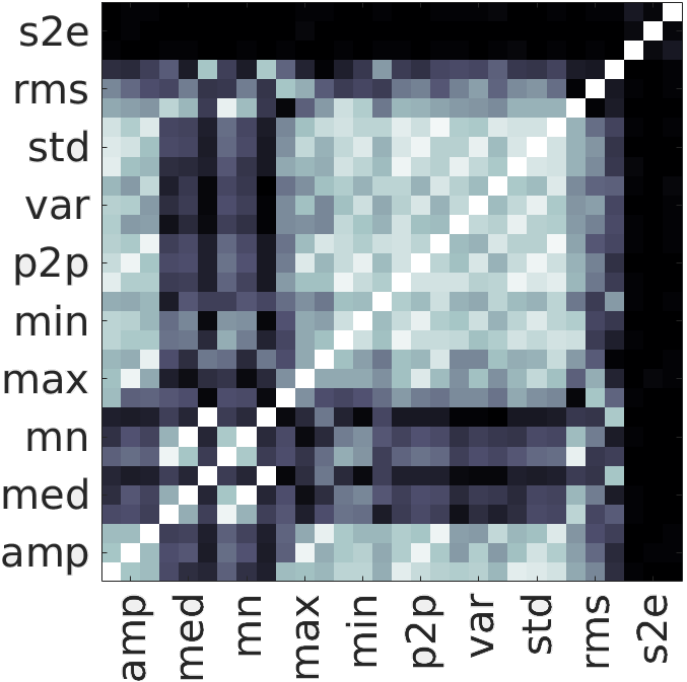}
  \label{fig:corrsRL}
 }
 \subfloat[Left leg.]{
  \includegraphics[width=0.37\columnwidth]{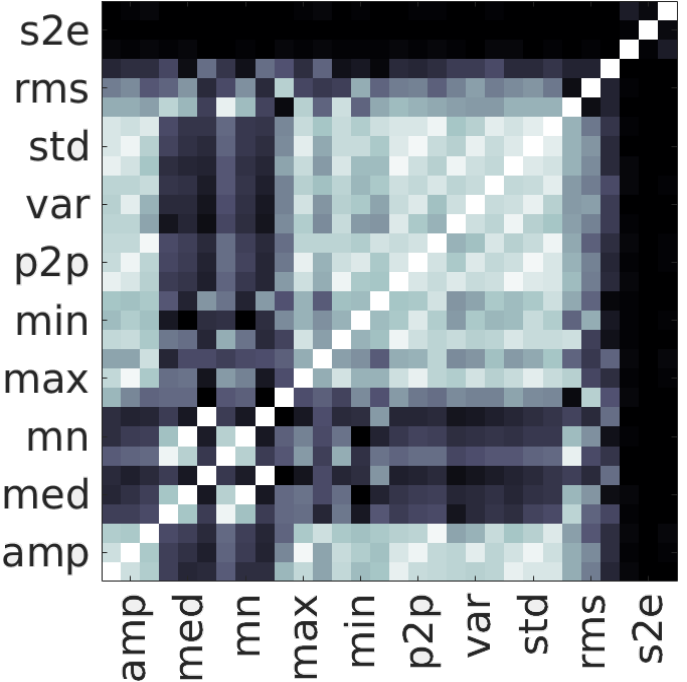}
  \label{fig:corrsLL}
 }
 \\
 \captionsetup[subfigure]{labelformat=empty}
 \subfloat[]{
  \includegraphics[width=\textwidth]{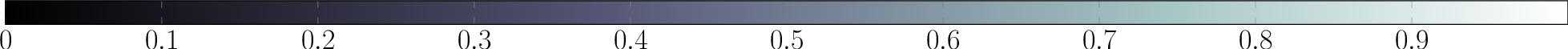} 
  \label{fig:colorbar}
 }
 \caption{Absolute value of the matrix of correlation of features for
each location considered. The lighter the color, the higher the
correlation.}
 \label{fig:corrs}
\end{figure*}

A preliminary study has been carried out in order to see correlation
between features, because some features might be highly correlated. To
see this,~\figref{corrs} shows the absolute value of the matrix of
correlations for all features and each location. As it can be
observed, correlation between features is similar among
locations. These matrices include all the subjects and activities. The
lighter the color, the higher the correlation. We can easily see that
there are only few features with low correlation between them; this
might help with the simplification of the system.

According to results in~\figref{corrs}, and results
in~\tblref{finalClusters}, we observe that i) only 1 feature is
necessary for the signal-type clustering, and ii) only few features
are not highly correlated.

Low correlated features in~\figref{corrs} are mostly those chosen by
the optimization process---median $med$ and mean $mn$ are well
represented. It is interesting to see that the $s2e$ values, although
they are not correlated with any other feature, are barely
selected. We can explain this fact as the information provided by
those signals is not enough to distinguish the signal type.

As can be seen, the selected features in~\tblref{finalClusters} match
with the expected ones regarding the distribution
in~\figref{histFeatures} too. So, we could reduce the number of
features to make more efficient the coarse-grained node localization
and the signal-type selection on sensing nodes.

For the sake of simplicity, this paper only presents a brief study to
show the feasibility of feature reduction. To do this we have reduced
the number of features down to 6 (median and mean for all axes), and
the study of the impact on energy savings due to data processing is
shown in~\tblref{energyPi}. In this table, it can be seen that using
$\#f=6$ consumption values of $FG_{\varepsilon}$ and
$NL_{\varepsilon}$ vary. Node localization using less features leads
to a higher consumption of $NL_{\varepsilon}$ as expected because
using less variables makes the classification harder and the trees are
larger (2467 nodes on average).

In this example, we showed that by further reducing the overhead of
computation, up to $14.7\%$ ($1 mJ$) extra energy savings were
reached. The overhead due to
computation---although still relatively small---can reach $14.2\%$ of total
energy consumption (see~\figref{energy}), and decrementing this value
in $1 mJ$ further helps to improve the battery life.

A more intelligent feature selection will draw better results,
reducing the maximum gap allowed of extra error. However feature
selection techniques require a deep knowledge of data. According to
the results drawn in previous figures, it seems that it is worthwhile
to calculate the trade-off between accuracy loss and energy
savings. We aim to tackle this issue in our future work from an energy
efficiency perspective.

\subsection{Comparison with on-node activity recognition alternatives}
\label{subsec:onNodeAR}

On the contrary to the problem discussed all along this work, there
might be applications where the motion data are not necessary to be
transmitted and activity recognition is preferred to be done on
sensing node and only the recognized activity is sent to the
back-end. The assumptions in these applications are totally different
from ours. These will benefit of a low power consumption mainly due to
the reduction of data transmission, but motion data will not be
available in the back-end for other purposes---such as visualization
on further computation.

On-node activity recognition can be implemented in different
scenarios: i) over uncompressed data or ii) over compressed
data. There will be three major processing modules in the architecture
of scenario (i): fine-grained feature generation, fine grained node
localization and fine grained-activity recognition. The architecture
of scenario (ii) would be similar to the one proposed for our adaptive
temporal compressed sensing methodology in \figref{metho}, but
substituting the \emph{Measurement Matrix} for a coarse-grained
activity recognition module. Both scenarios remove all the processing
modules in the \emph{Extreme-end unit}. In the following lines we show
the energy consumption and savings of these two alternatives. The
accuracy of these implementations will be commented as well as shown
in \tblref{energyLabelTX}.

On-body node localization and on-node activity recognition can be
implemented using \emph{Random Forest} algorithms on the sensing
node. Please, note the reader that $30$ features are
extracted/generated for further processing on-node, and so, henceforth
we assume that size and consumption of the trained Random Forest
algorithms for both processes---on-body node localization and on-node
activity recognition----are similar and both consume $0.5~mJ$ as seen
in \tblref{energyPi} (an extra consumption overhead of $0.5 mJ$ is
imposed by the on-node activity recognition).

Consumption due to data transmission is $0.1~mJ$ in both
scenarios. This is negligible compared with the consumption of our
approach due to only the label of the activity is sent to the back-end
once every 5 seconds (to be comparable with our results).

\begin{table}[]
\caption{Energy consumption, energy savings and accuracy for on-node AR scenarios.}
  \centering \resizebox{1.0\columnwidth}{!}{
\begin{tabular}{c|cccccccc}
\hline
On-node & \multirow{ 2}{*}{Location} & \multirow{ 2}{*}{$\sigma (mJ)$} & \multirow{ 2}{*}{$\pi (mJ)$} & \multirow{ 2}{*}{$\tau (mJ)$} & $\varepsilon_{Total}$ & $\varepsilon_{Tot.Sav.}$ & \multirow{ 2}{*}{$\alpha_{AR} (\%)$} \\
AR & & & & & (mJ) & (\%) & \\ \hline
 & T & 13.4 & 4.4  & 0.1 & 17.9 & 74.5 & 91.0\\ 
\emph{Coarse-} & RA & 13.4 & 3.4 &  0.1 & 16.9 & 76.0 & 84.1  \\                     
\emph{grained} & LA & 13.4 & 3.5 &  0.1 & 17.0  & 75.8 & 91.2 \\                      
 & RL & 13.4 & 5.1  &  0.1 & 18.6 & 73.5 & 87.0 \\                      
 & LL & 13.4 & 5.9  &  0.1 & 19.4 & 72.4 & 87.6\\ \hline               
\emph{Fine-} & \multirow{ 2}{*}{All} & \multirow{ 2}{*}{13.4} & \multirow{ 2}{*}{10.5} & \multirow{ 2}{*}{0.1} & \multirow{ 2}{*}{24.0} & \multirow{ 2}{*}{65.9} & \multirow{ 2}{*}{98.2} \\
\emph{grained} & & & & & & & \\ \hline
\end{tabular}
}
\label{tbl:energyLabelTX}
\end{table}

Scenario (i): if feature extraction is carried out over uncompressed
data, the feature extraction process will consume considerably more
energy. In this scenario the consumption is the same independently of
the localization. The average total energy saving increases $17.2$
points from $48.7\%$ to $65.9\%$ compared to our proposal. The average
accuracy, will be the one expected for the Baseline solution $98.2\%$
(see \tblref{fineLocation}).

Scenario (ii): the feature extraction module in this scenario consumes
the same than in ours. The average energy saving is $74.4\%$, which is
considerably higher than the ones achieved for our proposal and on-node
dine-grained AR. However, when we apply activity recognition over
sparse data (\ie~coarse-grained AR), we should expect a reduction in
the accuracy. \tblref{energyLabelTX} lists a comparison of accuracy of
AR ($\alpha_{AR}$). Accuracy drops down to $88.2\%$, and it is
worthwhile to mention that this level does violate our $5\%$ threshold
on accuracy decline (remember the reader that our adaptive temporal
compressed sensing methodology reaches $95.7\%$ average accuracy in
the back-end).

As aforementioned, the comparison of the methodology defended in this
paper with the two alternatives that implement on-node activity
recognition is not fair, as the assumptions and applications are
totally different. The achieved increments in energy saving are
accompanied by loosing the original motion data in the back-end, and a
possible reduction in the accuracy.



\section{Discussion}\label{sec:discussion}
\label{sec:discussion}

A variation of the proposed system can be developed without on-node
computation where the signal type detection and on-body node
localization are performed on the back-end and the feedback are
transmitted to the sensing node for adjustment of compression
ratio. The main benefit of this alternative approach is the
elimination of processing overhead on the local node. However, the
additional cost of constant transmission of feedback from back-end to
the sensing node must be taken into account. In addition, the
practical limitations that this alternative approach poses on the
system makes it less favorable in many real-world applications of
wearable activity recognition. The constant and real-time
dependability of this approach is often not affordable. There are two
reasons for this argument:
  
i) Assuming constant connectivity is too optimistic. Many wearable
devices rely on close-range, low-cost communication technologies such
as BLE, and are operated by humans in highly mobile settings. These
system often buffer the data, when out of range of the central node,
and transmit it once the central node become reachable.
  
ii) In practical settings, frequent sensor data/feedback transmission
is not favorable due to its large power consumption. Many of more
recent wearable technologies such smart-watches offer multiple means
of connectivity such as LTE, WiFi and BLE. Efficient applications
should aim to maximize their transmission on less costly links by
locally storing data until a reliable and cheap connection become
available. As a result, we argue that, with comparable overall energy
cost, the proposed approach is far more practical than the
aforementioned alternative because it does not need to transmit sensor
data (since it does not rely on external feedback).

In our methodology we stated a $5\%$ extra error as a quality
criterion proportional to the original baseline accuracy for activity
recognition (AR). The parameter $\epsilon_{th}$ in \eref{finalSols} is
defined to ensure an acceptable performance and avoid
over-minimization of sensing rates at the cost of end-result
accuracy. It can be viewed as a tuning parameter that can be adjusted
by an inference drawn from the domain knowledge (\eg~given the
accuracy of the baseline activity recognition algorithm, what is the
lowest acceptable accuracy that your application will consider
acceptable?). It is worthwhile to mention that larger thresholds will
result in more power optimization but will allow for significant
performance decline. On the other hand, excessively small thresholds
will not allow for significant data sensing optimization and therefore
are not desirable in energy stringent applications.



\section{Conclusion}\label{sec:conclusions}
\label{sec:conclusions}
The proposed methodology solves, through a novel adaptive temporal
compressed sensing technique, an important problem of the monitoring
devices in the paradigms of the MCC and RHM: battery consumption due
to excessive wireless transmissions. In this work we apply our
methodology in a physical activity recognition problem, but the system
is extensible to any other application of the IoT where power
efficiency is an obstacle. Utilizing metaheuristic optimization
techniques based on GE, our proposal achieves energy savings in
transmission of up to $81.2\%$, with negligible energy overhead in the
monitoring devices, which leads to global savings of up to $61.0\%$.

The proposed framework performs coarse-grained on-body sensor
localization and unsupervised clustering algorithms are employed to
autonomously reconfigure compressed sensing ratios ranging from
$42.8\%$ to $77.4\%$ on average for different on-body node
localization. With this approach, we achieve significant accuracy
levels between $87.7\%$ and $90.2\%$ when performing fine-grained
activity recognition in the back-end computing unit (e.g. a smartphone
or a data center).

The proposed optimized adaptive temporal compressed sensing
methodology has a bounded error of $5\%$ over the baseline, and
reaches higher energy savings when compared with the state-of-the-art:
a naive compressed sensing approach with invariable compression
ratio. A coarse-grained on-body sensor localization---based on
a \emph{Random Forest}---is performed and it has been shown that the
error in node localization is not propagated to the fine-grained
activity recognition in the back-end unit.

An additional study about the reduction of the number of features has
been carried out. Preliminary results showed that, after our feature
optimization, we can achieve up to $14.7\%$ energy savings in
computation, which further contributes to an extended battery life.



\section*{Acknowledgement}
This work was supported in part by the EU (FEDER) and the Spanish
Ministry of Economy and Competitiveness under Research Grants
TIN2015-65277-R and TEC2012-33892 and the National Science Foundation
under grant CNS-1566359. Any opinions, findings, conclusions, or
recommendations expressed in this material are those of the authors
and do not necessarily reflect the views of the funding organizations.



\bibliographystyle{IEEEtran}
\bibliography{bibliography}

\begin{thebibliography}{10}
\providecommand{\url}[1]{#1}
\csname url@samestyle\endcsname
\providecommand{\newblock}{\relax}
\providecommand{\bibinfo}[2]{#2}
\providecommand{\BIBentrySTDinterwordspacing}{\spaceskip=0pt\relax}
\providecommand{\BIBentryALTinterwordstretchfactor}{4}
\providecommand{\BIBentryALTinterwordspacing}{\spaceskip=\fontdimen2\font plus
\BIBentryALTinterwordstretchfactor\fontdimen3\font minus
  \fontdimen4\font\relax}
\providecommand{\BIBforeignlanguage}[2]{{%
\expandafter\ifx\csname l@#1\endcsname\relax
\typeout{** WARNING: IEEEtran.bst: No hyphenation pattern has been}%
\typeout{** loaded for the language `#1'. Using the pattern for}%
\typeout{** the default language instead.}%
\else
\language=\csname l@#1\endcsname
\fi
#2}}
\providecommand{\BIBdecl}{\relax}
\BIBdecl

\bibitem{zheng2014unobtrusive}
Y.-L. Zheng, X.-R. Ding, and et~al., ``Unobtrusive sensing and wearable devices
  for health informatics,'' \emph{Biomedical Engineering, IEEE Transactions
  on}, vol.~61, no.~5, pp. 1538--1554, 2014.

\bibitem{kugler2011mobile}
P.~Kugler, D.~Schuldhaus, U.~Jensen, and B.~Eskofier, ``Mobile recording system
  for sport applications,'' in \emph{Proceedings of IACSS 2011, Liverpool},
  2011, pp. 67--70.

\bibitem{alemdar2010wireless}
H.~Alemdar and C.~Ersoy, ``Wireless sensor networks for healthcare: A survey,''
  \emph{Computer Networks}, vol.~54, no.~15, pp. 2688--2710, 2010.

\bibitem{steele2000quantitating}
B.~G. Steele, L.~Holt, B.~Belza, S.~Ferris, S.~Lakshminaryan, and D.~M.
  Buchner, ``Quantitating physical activity in copd using a triaxial
  accelerometer,'' \emph{CHEST Journal}, vol. 117, no.~5, pp. 1359--1367, 2000.

\bibitem{klucken2013unbiased}
J.~Klucken, J.~Barth, and et~al., ``Unbiased and mobile gait analysis detects
  motor impairment in parkinson's disease,'' \emph{PloS one}, vol.~8, no.~2, p.
  e56956, 2013.

\bibitem{rashidi2013survey}
P.~Rashidi and A.~Mihailidis, ``A survey on ambient-assisted living tools for
  older adults,'' \emph{IEEE journal of biomedical and health informatics},
  vol.~17, no.~3, pp. 579--590, 2013.

\bibitem{saeedi2014autolocate}
R.~Saeedi and H.~Ghasemzadeh, ``Autolocate: A machine learning approach for
  automatic localization of wearable sensors in smart health applications,''
  Abstract (Academic Showcase at Washington State University, Pullman, WA),
  2014.

\bibitem{saeedi2014toward}
R.~Saeedi, J.~Purath, K.~Venkatasubramanian, and H.~Ghasemzadeh, ``Toward
  seamless wearable sensing: Automatic on-body sensor localization for physical
  activity monitoring,'' in \emph{Engineering in Medicine and Biology Society
  (EMBC), 2014 36th Annual International Conference of the IEEE}.\hskip 1em
  plus 0.5em minus 0.4em\relax IEEE, 2014, pp. 5385--5388.

\bibitem{bhattacharya2016smart}
S.~Bhattacharya and N.~D. Lane, ``From smart to deep: Robust activity
  recognition on smartwatches using deep learning,'' in \emph{2016 IEEE PerCom
  Workshops}.\hskip 1em plus 0.5em minus 0.4em\relax IEEE, 2016, pp. 1--6.

\bibitem{maurer2006location}
U.~Maurer, A.~Rowe, A.~Smailagic, and D.~Siewiorek, ``Location and activity
  recognition using ewatch: A wearable sensor platform,'' in \emph{Ambient
  Intelligence in Everyday Life}.\hskip 1em plus 0.5em minus 0.4em\relax
  Springer, 2006, pp. 86--102.

\bibitem{patterson2005fine}
D.~J. Patterson, D.~Fox, H.~Kautz, and M.~Philipose, ``Fine-grained activity
  recognition by aggregating abstract object usage,'' in \emph{IEEE
  ISWC'05}.\hskip 1em plus 0.5em minus 0.4em\relax IEEE, 2005, pp. 44--51.

\bibitem{choudhury2008mobile}
T.~Choudhury, G.~Borriello, and et~al., ``The mobile sensing platform: An
  embedded activity recognition system,'' \emph{IEEE Pervasive Computing},
  vol.~7, no.~2, pp. 32--41, 2008.

\bibitem{akimura2012compressed}
D.~Akimura, Y.~Kawahara, and T.~Asami, ``Compressed sensing method for human
  activity sensing using mobile phone accelerometers,'' in \emph{9th INSS,
  2012}.\hskip 1em plus 0.5em minus 0.4em\relax IEEE, 2012, pp. 1--4.

\bibitem{khan2013towards}
A.~N. Khan, M.~Kiah, and et~al., ``Towards secure mobile cloud computing: A
  survey,'' \emph{FGCS}, vol.~29, no.~5, pp. 1278--1299, 2013.

\bibitem{santambrogio2015power}
M.~Santambrogio and et~al., ``Power-awareness and smart-resource management in
  embedded computing systems,'' in \emph{CODES+ISSS}.\hskip 1em plus 0.5em
  minus 0.4em\relax IEEE Press, 2015, pp. 94--103.

\bibitem{khalifa2017harke}
S.~Khalifa, G.~Lan, M.~Hassan, A.~Seneviratne, and S.~K. Das, ``Harke: Human
  activity recognition from kinetic energy harvesting data in wearable
  devices,'' \emph{IEEE Transactions on Mobile Computing}, vol.~17, no.~6, pp.
  1353--1368, June 2018.

\bibitem{lan2017capsense}
G.~Lan, D.~Ma, W.~Xu, M.~Hassan, and W.~Hu, ``Capsense: Capacitor-based
  activity sensing for kinetic energy harvesting powered wearable devices,'' in
  \emph{Proceedings of the 14th EAI International Conference on Mobile and
  Ubiquitous Systems (MobiQuitous'17)}, 2017, pp. 110--119.

\bibitem{mamaghanian2011compressed}
H.~Mamaghanian and et~al., ``Compressed sensing for real-time energy-efficient
  ecg compression on wireless body sensor nodes,'' \emph{IEEE Transactions on
  Biomedical Engineering}, vol.~58, no.~9, pp. 2456--2466, 2011.

\bibitem{chen2014compressed}
S.-W. Chen and S.-C. Chao, ``Compressed sensing technology-based spectral
  estimation of heart rate variability using the integral pulse frequency
  modulation model,'' \emph{IEEE journal of biomedical and health informatics},
  vol.~18, no.~3, pp. 1081--1090, 2014.

\bibitem{liu2015compressed}
Y.~Liu, M.~De~Vos, and S.~Van~Huffel, ``Compressed sensing of multichannel eeg
  signals: the simultaneous cosparsity and low-rank optimization,'' \emph{IEEE
  Transactions on Biomedical Engineering}, vol.~62, no.~8, pp. 2055--2061,
  2015.

\bibitem{baheti2009ultra}
P.~K. Baheti and H.~Garudadri, ``An ultra low power pulse oximeter sensor based
  on compressed sensing,'' in \emph{Wearable and Implantable Body Sensor
  Networks, 2009. BSN 2009. Sixth International Workshop on}.\hskip 1em plus
  0.5em minus 0.4em\relax IEEE, 2009, pp. 144--148.

\bibitem{fallahzadeh2017adaptive}
R.~Fallahzadeh, J.~P. Ortiz, and H.~Ghasemzadeh, ``Adaptive compressed sensing
  at the fingertip of internet-of-things sensors: An ultra-low power activity
  recognition,'' in \emph{2017 Design, Automation \& Test in Europe Conference
  \& Exhibition (DATE)}.\hskip 1em plus 0.5em minus 0.4em\relax IEEE, 2017, pp.
  996--1001.

\bibitem{razzaque2014energy}
M.~A. Razzaque and et~al., ``Energy-efficient sensing in wireless sensor
  networks using compressed sensing,'' \emph{Sensors}, vol.~14, no.~2, pp.
  2822--2859, 2014.

\bibitem{candes2008introduction}
E.~J. Cand{\`e}s and M.~B. Wakin, ``An introduction to compressive sampling,''
  \emph{IEEE signal processing magazine}, vol.~25, no.~2, pp. 21--30, 2008.

\bibitem{eldar2012compressed}
Y.~C. Eldar and G.~Kutyniok, \emph{Compressed sensing: theory and
  applications}.\hskip 1em plus 0.5em minus 0.4em\relax Cambridge University
  Press, 2012.

\bibitem{donoho2006compressed}
D.~L. Donoho, ``Compressed sensing,'' \emph{IEEE Transactions on information
  theory}, vol.~52, no.~4, pp. 1289--1306, 2006.

\bibitem{candes2006robust}
E.~J. Cand{\`e}s, J.~Romberg, and T.~Tao, ``Robust uncertainty principles:
  Exact signal reconstruction from highly incomplete frequency information,''
  \emph{IEEE Transactions on information theory}, vol.~52, no.~2, pp. 489--509,
  2006.

\bibitem{natarajan1995sparse}
B.~K. Natarajan, ``Sparse approximate solutions to linear systems,'' \emph{SIAM
  journal on computing}, vol.~24, no.~2, pp. 227--234, 1995.

\bibitem{yuan2013adaptive}
X.~Yuan, J.~Yang, P.~Llull, X.~Liao, G.~Sapiro, D.~J. Brady, and L.~Carin,
  ``Adaptive temporal compressive sensing for video,'' in \emph{Image
  Processing (ICIP), 2013 20th IEEE International Conference On}.\hskip 1em
  plus 0.5em minus 0.4em\relax IEEE, 2013, pp. 14--18.

\bibitem{chiu2015jice}
S.-Y. Chiu, H.~H. Nguyen, R.~Tan, D.~K. Yau, and D.~Jung, ``Jice: Joint data
  compression and encryption for wireless energy auditing networks,'' in
  \emph{Sensing, Communication, and Networking (SECON), 2015 12th Annual IEEE
  International Conference on}.\hskip 1em plus 0.5em minus 0.4em\relax IEEE,
  2015, pp. 453--461.

\bibitem{duarte2008single}
M.~F. Duarte, M.~A. Davenport, D.~Takbar, J.~N. Laska, T.~Sun, K.~F. Kelly, and
  R.~G. Baraniuk, ``Single-pixel imaging via compressive sampling,'' \emph{IEEE
  signal processing magazine}, vol.~25, no.~2, pp. 83--91, 2008.

\bibitem{constantin2012ultra}
J.~Constantin, A.~Dogan, O.~Andersson, P.~Meinerzhagen, J.~Rodrigues,
  D.~Atienza, and A.~Burg, ``An ultra-low-power application-specific processor
  with sub-vt memories for compressed sensing,'' in \emph{IFIP/IEEE
  International Conference on Very Large Scale Integration-System on a
  Chip}.\hskip 1em plus 0.5em minus 0.4em\relax Springer, 2012, pp. 88--106.

\bibitem{yang2010energy}
S.~Yang and M.~Gerla, ``Energy-efficient accelerometer data transfer for human
  body movement studies,'' in \emph{IEEE SUTC 2010}.\hskip 1em plus 0.5em minus
  0.4em\relax IEEE, 2010, pp. 304--311.

\bibitem{ryan1998grammatical}
C.~Ryan and M.~O’Neill, ``Grammatical evolution: A steady state approach,''
  \emph{Late Breaking Papers, Genetic Programming}, vol. 1998, pp. 180--185,
  1998.

\bibitem{pagan2016grammatical}
J.~Pag{\'a}n, J.~L. Risco-Mart{\'\i}n, J.~M. Moya, and J.~L. Ayala,
  ``Grammatical evolutionary techniques for prompt migraine prediction,'' in
  \emph{Proceedings of the 2016 on Genetic and Evolutionary Computation
  Conference}.\hskip 1em plus 0.5em minus 0.4em\relax ACM, 2016, pp. 973--980.

\bibitem{deb2002fast}
K.~Deb, A.~Pratap, S.~Agarwal, and T.~Meyarivan, ``A fast and elitist
  multiobjective genetic algorithm: Nsga-ii,'' \emph{IEEE transactions on
  evolutionary computation}, vol.~6, no.~2, pp. 182--197, 2002.

\bibitem{colmenar2011multi}
J.~M. Colmenar, J.~L. Risco-Martin, D.~Atienza, and J.~I. Hidalgo,
  ``Multi-objective optimization of dynamic memory managers using grammatical
  evolution,'' in \emph{Proceedings of the 13th GECCO}.\hskip 1em plus 0.5em
  minus 0.4em\relax ACM, 2011, pp. 1819--1826.

\bibitem{davies1979cluster}
D.~L. Davies and et~al., ``A cluster separation measure,'' \emph{IEEE TPAMI},
  no.~2, pp. 224--227, 1979.

\bibitem{HERO}
HERO, ``Heuristic optimization library,''
  \url{https://github.com/jlrisco/hero}, 2016.

\bibitem{candes2005l1}
E.~Candes and J.~Romberg, ``l1-magic: Recovery of sparse signals via convex
  programming,'' \emph{URL: www. acm. caltech. edu/l1magic/downloads/l1magic.
  pdf}, vol.~4, p.~14, 2005.

\bibitem{hall2009weka}
M.~Hall, E.~Frank, and et~al., ``The weka data mining software: an update,''
  \emph{ACM SIGKDD}, vol.~11, no.~1, pp. 10--18, 2009.

\bibitem{breiman2001random}
L.~Breiman, ``Random forests,'' \emph{Machine learning}, vol.~45, no.~1, pp.
  5--32, 2001.

\bibitem{barshan2014recognizing}
B.~Barshan and et~al., ``Recognizing daily and sports activities in two open
  source machine learning environments using body-worn sensor units,''
  \emph{The Computer Journal}, vol.~57, no.~11, pp. 1649--1667, 2014.

\bibitem{ghasemzadeh2015power}
H.~Ghasemzadeh, N.~Amini, and et~al., ``Power-aware computing in wearable
  sensor networks: An optimal feature selection,'' \emph{IEEE Transactions on
  Mobile Computing}, vol.~14, no.~4, pp. 800--812, 2015.

\end{thebibliography}


%

\begin{IEEEbiography}[{\includegraphics[width=1in,height=1.25in,clip,keepaspectratio]{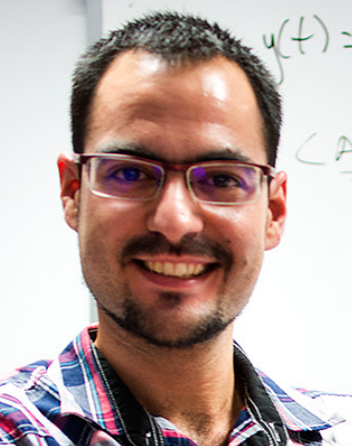}}]{Josué Pagán}
is a PhD candidate at the Complutense University of Madrid and works
as a researcher at the Technical University of Madrid. His work
focuses on the development of robust methodologies for information
acquisition in biophysical and critical scenarios. He has worked
developing models for prompt prediction and classification of
neurological diseases. In 2016 he workED as researcher at the Embedded
Pervasive Systems Lab at Washington State University under the
supervision of Prof. Hassan Ghasemzadeh. In 2015 he worked as
researcher at the Pattern Recognition Lab. at Friedrich Alexander
University. He earned his MSc at the Technical University of Madrid in
2013 with honors. He is a student member of the IEEE.
\end{IEEEbiography}

\vspace{-1cm}

\begin{IEEEbiography}[{\includegraphics[width=1in,height=1.25in,clip,keepaspectratio]{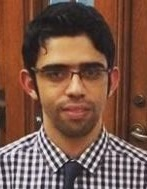}}]{Ramin Fallahzadeh}
received his B.S. degree in computer engineering from Sharif
University of Technology, Tehran, Iran in 2014. Currently, he is a PhD
candidate in computer science at Washington State University. His
current research interests include smart-health, pervasive computing,
machine learning, and wireless sensor networks. The focus of his
research is on algorithm design and power optimization of networked
wearable sensors with applications in healthcare. He is a student
member of the IEEE.
\end{IEEEbiography}

\vspace{-1cm}

\begin{IEEEbiography}[{\includegraphics[width=1in,height=1.25in,clip,keepaspectratio]{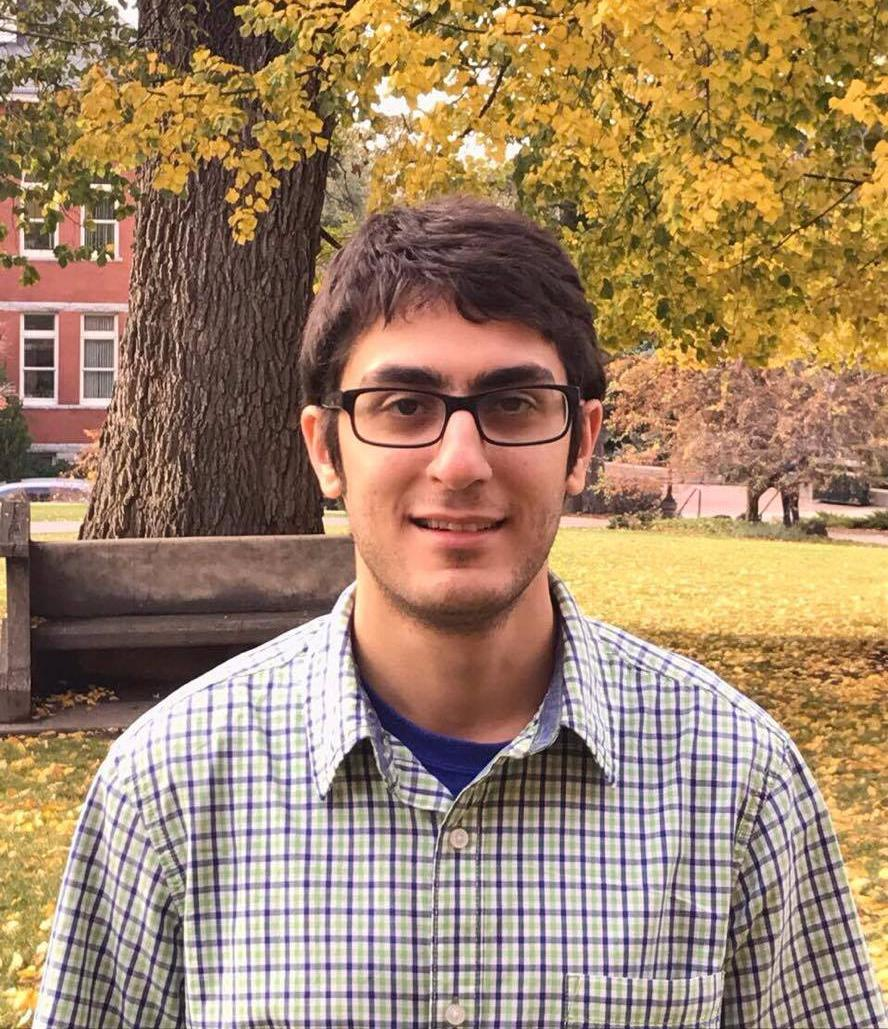}}]{Mahdi Pedram} 
received the B.S. degree in computer engineering from Amirkabir
University of Technology, Tehran, Iran in 2014. Currently he is a PhD
student with Advanced Graduate Standing status in computer engineering
at Washington State University. He has been working as a research
assistant in the Embedded \& Pervasive Systems Lab since January,
2016. His research interests include body sensor networks, pervasive
computing in healthcare and embedded system design. He is a student
member of the IEEE.
\end{IEEEbiography}

\vspace{-1cm}

\begin{IEEEbiography}[{\includegraphics[width=1in,height=1.25in,clip,keepaspectratio]{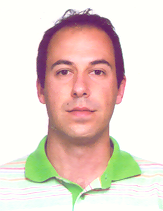}}]{José L. Risco-Martín}
  is an Associate Professor at the Computer Architecture and Automation
  Department of Complutense University of Madrid, Spain. His
  research interests focus on design methodologies for integrated
  systems and high-performance embedded systems, including new modeling
  frameworks to explore thermal management techniques for
  Multi-Processor System-on-Chip, novel architectures for logic and
  memories in forthcoming nano-scale electronics, dynamic memory
  management and memory hierarchy optimization for embedded systems,
  Networks-on-Chip interconnection design, and low-power design of
  embedded systems.
\end{IEEEbiography}

\vspace{-1cm}

\begin{IEEEbiography}[{\includegraphics[width=1in,height=1.25in,clip,keepaspectratio]{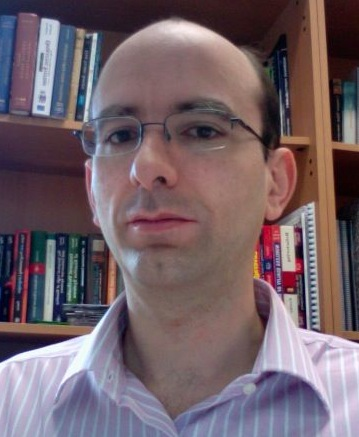}}]{José M. Moya}
is currently an Associate Professor in the Department of Electronic
Engineering, Technical University of Madrid. He received his MSc and
PhD degrees in Telecommunication Engineering from the Technical
University of Madrid, Spain, in 1999 and 2003, respectively. He is
member of the HiPEAC European Network of Excellence. He has served as
TPC member of many conferences, including DATE, IEEE ICC, IEEE
MASS~\etc~He has participated in a large number of national research
projects and bilateral projects with industry, in the fields of
embedded system design and optimization, and security optimization of
embedded systems and distributed embedded systems. His current
research interests focus on proactive and reactive thermal-aware
optimization of data centers, and design techniques and tools for
energy-efficient compute-intensive embedded applications.
\end{IEEEbiography}

\vspace{-1cm}

\begin{IEEEbiography}[{\includegraphics[width=1in,height=1.25in,clip,keepaspectratio]{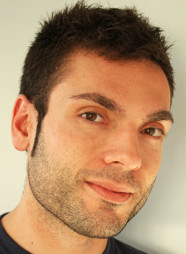}}]{José L. Ayala}
is currently an Associate Professor at the Computer Architecture and
Automation Department of Complutense University of Madrid. He received
his MSc and PhD degrees in Telecommunication Engineering from the
Technical University of Madrid in 2001 and 2005, respectively. He is
member of the HiPEAC European Network of Excellence, IEEE, ACM, IFIP
10.5 and the CEDA. He has organized several international events as
General and Program Chair, such as VLSI-SoC. He has served as TPC
member of many conferences, including DATE, DAC, ICCAD~\etc~. His
current research interests focus on thermal and energy aware design,
design of embedded processors, thermal estimation, 3D integration,
health monitoring and wireless sensor networks. He is a senior member
of the IEEE.
\end{IEEEbiography}

\vspace{-1cm}

\begin{IEEEbiography}[{\includegraphics[width=1in,height=1.25in,clip,keepaspectratio]{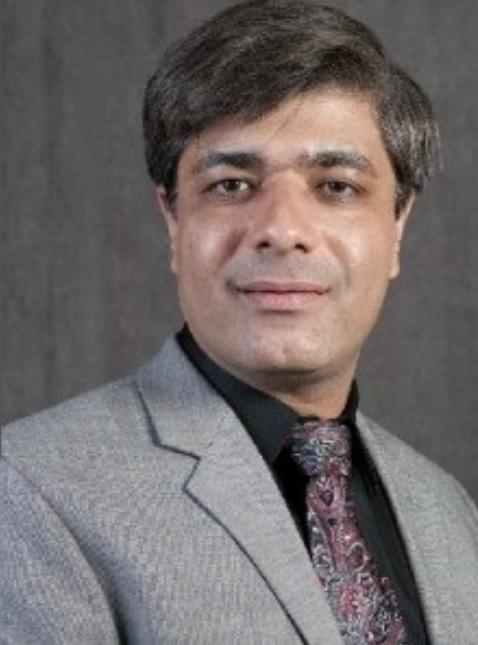}}]{Hassan Ghasemzadeh}
received the B.Sc. degree from Sharif University of Technology,
Tehran, Iran, the M.Sc. from University of Tehran, Tehran, Iran, and
the Ph.D. from the University of Texas at Dallas, Richardson, TX, in
1998, 2001, and 2010 respectively, all in Computer Engineering. He was
on the faculty of Azad University from 2003-2006 where he served as
Founding Chair of Computer Science and Engineering Department at
Damavand branch, Tehran, Iran. He spent the academic year 2010- 2011
as a Postdoctoral Fellow at the West Wireless Health Institute, La
Jolla, CA. He was a Research Manager at UCLA Wireless Health Institute
2011-2013. Currently, he is Assistant Professor in Electrical
Engineering and Computer Science at Washington State University,
Pullman, WA. The focus of his research is on algorithm design and
system level optimization of embedded and pervasive systems with
applications in healthcare and wellness. He is a senior member of the
IEEE.
\end{IEEEbiography}







\end{document}